\documentclass[11pt]{article}
\usepackage{tabularx}
\usepackage[utf8]{inputenc}
\usepackage{hyperref}
\usepackage{setspace}
\usepackage{palatino}
\usepackage{graphicx}
\usepackage{float}
\usepackage{titling} %
\usepackage{multirow}
\usepackage{lscape}
\usepackage{amsmath}
\usepackage{amssymb}
\usepackage{subcaption}
\usepackage{booktabs}

\usepackage[a4paper, total={6in, 9.5in}]{geometry}
\fontfamily{ppl}\selectfont

\usepackage[american]{babel}

\providecommand{\JournalTitle}[1]{#1}

\title{Return of the solo author: The changing division of labor in science in the age of generative AI}
\author{
Akira Matsui\\
Center for Computational Social Science, Kobe University, Kobe, Hyogo, Japan\\
\href{mailto:amatsui@rieb.kobe-u.ac.jp}{\texttt{amatsui@rieb.kobe-u.ac.jp}}
}
\date{\today}

\begin{document}

\maketitle

\begin{abstract}
Modern science has experienced a long shift from individual work to team production. Generative artificial intelligence (AI) might appear to extend this trajectory by lowering research costs and enabling larger-scale collaboration. Yet if tasks once performed by coauthors can be delegated to AI, the same technology may also weaken the need for collaboration in parts of the research process. Here, we examine this tension by moving beyond average team size and focusing on the solo-authored tail of the author-count distribution. Analyzing over 300 million works across 26 fields, we find that the decades-long decline in solo authorship halted and partially reversed with ChatGPT's public release in late 2022. We also reveal that this is an uneven phenomenon: it is strongest in fields where coauthors' work is more readily replaceable, and weak or absent in fields that depend on physical collaboration. At the individual level, the recovery is not explained by the entry of new researchers or by changes in field composition. Instead, the break appears among authors who had written only with others, including those with no prior solo publications, and among long-established authors as well as newcomers. Their solo papers stay close to their own coauthored work while narrowing in scope and shifting toward computational topics. Because a solo paper is work without credited human coauthors, this study offers an empirical probe of how generative AI can substitute for scientific labor, and evidence of a reconfiguration of cognitive labor within papers rather than of team size.
\end{abstract}


\bigskip

The division of labor is one of the most basic mechanisms for raising productivity.
As Adam Smith described with the pin factory, dividing a single task into several specialized steps, each carried out by a different person, dramatically increases total output.
Science is no exception.
What science has divided, however, is not the labor of producing things but cognitive labor~\cite{Jones2009}.
By forming research teams, researchers can divide their cognitive workload~\cite{wuchty2007,haeussler2020division},
a necessity in scientific research that has grown ever more complex and specialized. As instruments, data, and collaborations have grown in scale and complexity, the cognitive labor behind a single paper has been divided among ever more specialized contributors~\cite{Fortunato2018,wang2021sciscibook}.
The well-known rise of team science is reflected in the decades-long increase in the average number of authors per paper~\cite{Leahey2016, Thelwall2022, Guimera2005, Milojevic2014}, and such teams can result in impactful outcomes~\cite{wuchty2007, wu2019large}.
If team science reflects a deepening division of cognitive labor, how does generative AI affect that division?
There are two views, and they make opposite predictions.
The first is the \emph{acceleration view}.
LLMs have been argued to lower the cost of producing papers, increasing the number of coauthors and encouraging gift authorship~\cite{cunningham2025}.
On this view, generative AI extends the division of labor: more contributors join each paper, and the upper tail of the author-count distribution grows.
The second is the \emph{substitution view}, which focuses on the tasks allocated to individual contributors.
Research tasks such as writing, data preparation, and statistical analysis have conventionally been divided among human coauthors~\cite{xu2022}.
LLMs now perform many of these tasks, from writing~\cite{Noy2023} to coding and data analysis~\cite{Cui2026, Korinek2023}.
On this view, tasks in the division of labor shift from humans to LLMs, and researchers can complete work that once required collaborators on their own.
This would imply an increase in solo authorship, that is, a thickening of the left tail of the author-count distribution.
The two views therefore predict opposite effects of generative AI on team size.
Distinguishing them is not straightforward, because the average number of authors can rise under both: the acceleration view is consistent with a rising mean, and the mean is insensitive to changes in the left tail.
To settle this tension, this study focuses on the left tail of the author-count distribution, which a measure of central tendency cannot capture.
Using the full OpenAlex data (over 300 million works, of which 126 million are peer-reviewed papers; 26 fields, 1990--2025), we ask whether the decades-long decline in solo authorship has reversed, whether it is explained by a changing mix of authors, and whether the pattern is consistent with LLMs taking over the role of coauthors.
Our goal is to turn solo authorship into an observable probe of which parts of scientific labor generative AI can replace.

\section*{Results}
\begin{figure*}[t]
  \centering
  \includegraphics[width=\textwidth]{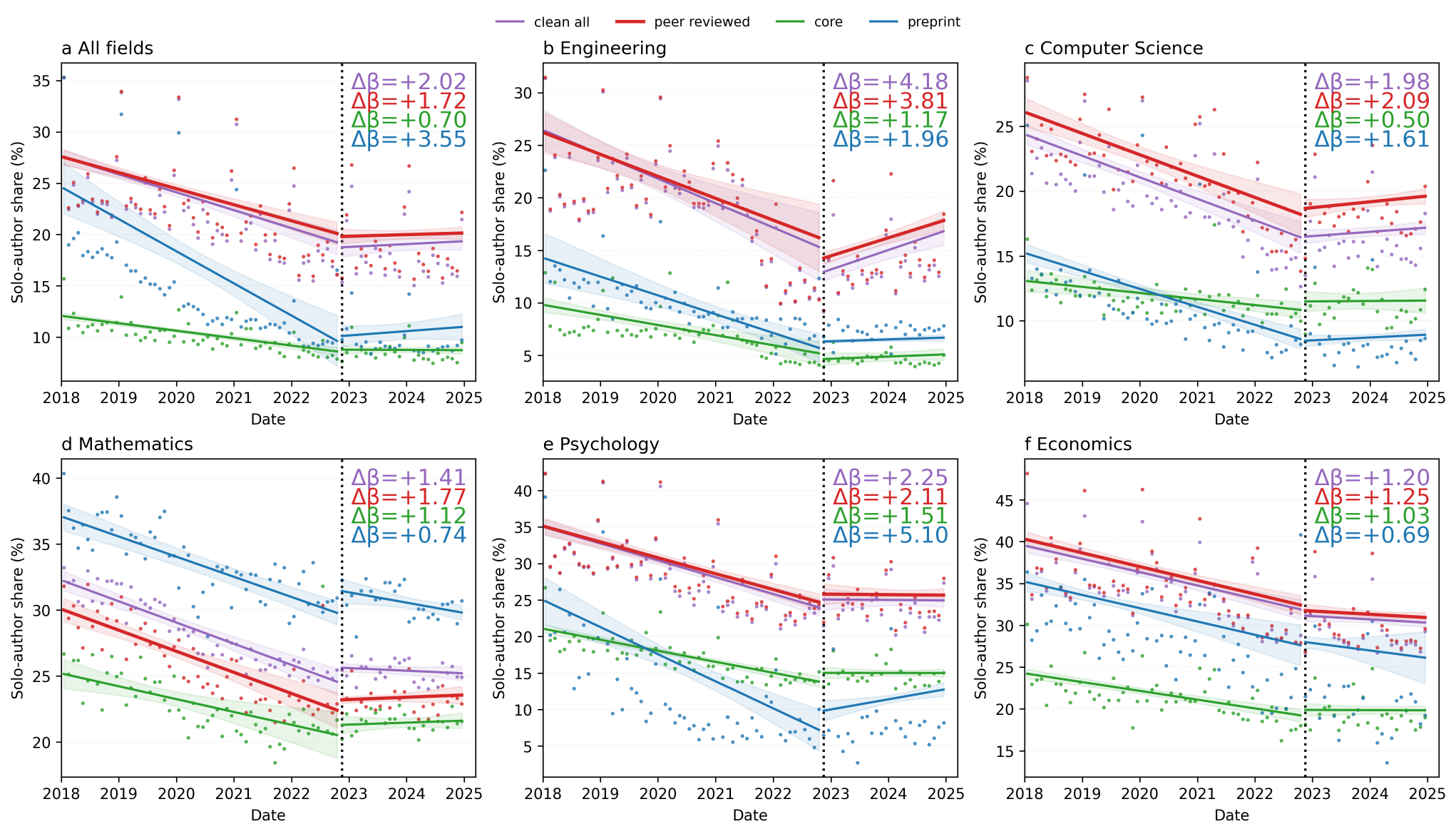}
  \caption[Piecewise-linear trend break in the monthly solo-author share]{%
  \textbf{Piecewise-linear break in the monthly solo-author share at the public release of ChatGPT.}
  Panels show the monthly share of solo-authored papers from January 2018 through December 2024 for all fields pooled and for five focal fields: Engineering, Computer Science, Mathematics, Psychology, and Economics.
  Colors distinguish four coverage variants, with \emph{peer reviewed} used as the main specification and the other variants shown as robustness checks.
  Dots are monthly observations; the vertical dotted line marks November 2022, when ChatGPT was publicly released.
  Separate weighted linear fits are estimated before and after this date, and the shaded bands show 95\% confidence intervals for the fitted trends; the year 2025 is excluded because of indexing lag.
  The reported quantity, $\Delta\beta=\beta_{\mathrm{post}}-\beta_{\mathrm{pre}}$, is the change in the annual slope of the solo-authorship share.
  Positive values indicate that the pre-release decline in solo authorship flattened or reversed after the release.
  }
  \label{fig:main}
\end{figure*}
\subsection*{The decline in solo authorship halts or reverses with ChatGPT and is observed broadly}
The share of solo-authored papers had declined for a long time~\cite{Leahey2016, Thelwall2022, Guimera2005, Milojevic2014, cunningham2025,  praus2025note}.
As team science advanced, solo-authored papers gradually disappeared across fields.
Yet when we track the monthly share of solo authorship and fit separate linear trends before and after the public release of ChatGPT, the slope of this decline flattens or reverses at the release (Fig.~1).
This halt in the downward trend has persisted from late 2022 to the present, and it holds across all four coverage variants.
For empirical purposes, monthly analyses split the series at the release date (November 2022), and yearly analyses use 2022 as the transition year, with its exact assignment to the pre or post window stated for each estimator (Materials and Methods; SI, Section~B).
This cutoff is a dating convention; it does not claim that effects appear immediately in print. Because LLM adoption is gradual and unobservable, we anchor the analysis at the only sharp timestamp common to all fields and authors.
This approach parallels economic research on the impact of COVID-19, which uses the observable onset of the pandemic as a common temporal reference even though behavioral responses emerged gradually, differed across actors, and sometimes preceded formal policy interventions~\cite{GoolsbeeSyverson2021,chetty2024economic}.
We therefore estimate the trend in the solo-authorship rate for each year and compute the change in the estimated slope around each cutoff year, denoted by $c$, as $\Delta\beta_c$ (SI, Section~B).
We find that the reversal is not a one-off event right after the release: the slope change remains positive in the years that follow (SI Figs.~S1 and S4).
This indicates that a long-standing trend has changed.
Publication lags blur the timing of this break rather than dating it sharply: peer-reviewed papers published in early 2023 were largely written before the release.
Consistent with lags of this kind, the rebound is largest for preprints, where the lag between writing and appearance is shortest.
The break also survives excluding the transition year from the annual estimation windows, with the monthly point estimate essentially unchanged though less precisely estimated, and it appears as the abrupt halt of an accelerating decline in an event-study specification (SI, Section~M).
Moreover, the break is not specific to our tail-based metric.
Applying the same pre/post design to the outcome emphasized by prior work, the mean number of authors per paper, reveals a companion break: its decades-long rise flattens or turns downward after late 2022 in most fields and coverage variants, although less uniformly than the solo-share reversal (SI Fig.~S2).
In 23 of the 26 fields, the point estimate of the trend break turns positive under the main peer-reviewed specification, and in 22 of them the change is significant at $p<0.05$ (SI Fig.~S3, $c=2022$ column).
The three exceptions are Chemistry, Arts and Humanities, and Physics and Astronomy; only in Arts and Humanities does the decline remain statistically significant, while Chemistry and Physics and Astronomy are essentially flat. We return to these non-reversing fields below.
The recovery of solo authorship therefore occurs broadly and at the same time.
The cutoff-based estimates make this visible across fields and filter variants (SI Fig.~S1).
It is also present within a balanced panel of venues observed continuously over 2018--2024, which carry 72--77\% of peer-reviewed papers, so it is not driven solely by sources entering or leaving the index, although within these venues the decline decelerates sharply rather than turns positive (SI, Section~L).
Positive slope changes appear at earlier cutoffs as well, but they reflect a deceleration of the long decline, which remained ongoing; only at $c=2022$ does the post-window trend itself turn positive in a broad set of fields (11 of 26, versus at most 3 at any earlier cutoff).
This timing is also hard to explain with a pandemic-rebound account, in which the recovery would merely reflect solo shares reverting to trend after the expansion of large-team, COVID-related publishing (SI, Section~D).
This phenomenon, however, displays field-level heterogeneity: some fields reverse strongly, while others barely move. Under the main peer-reviewed specification, the trend break at the 2022 cutoff is largest in Engineering ($+2.5$~pp/yr) and Business, Management and Accounting ($+1.9$), followed by a tight cluster around $+1.7$ comprising Mathematics, Decision Sciences, Psychology, and Computer Science, while it is negative in Arts and Humanities ($-0.5$) and essentially flat in Chemistry ($-0.1$) and Physics and Astronomy ($-0.0$) (SI Fig.~S3, $c=2022$ column). This ordering is broadly robust to the window definition (SI, Section~E).
Engineering, however, is a special case: its author-level solo probability
does not move, and its field-level break attenuates from $+2.47$ to
$+0.65$~pp/yr within continuously observed venues, so the aggregate
reversal there operates through composition rather than through
within-author switching (SI, Sections~H and~L).
The reversal is therefore a broad phenomenon, but its size differs clearly and consistently across fields.
One possible interpretation of this heterogeneity is the substitutability of coauthors, as observed in labor markets~\cite{Eloundou2024, felten2023heterogeneity}. In fields in which the work previously done by coauthors centers on tasks that machines can readily take over, such as writing, coding, and statistical support, the reversal tends to be stronger, whereas fields that depend on laboratory work and instrument operation reverse weakly or not at all. The fields that stay essentially flat under both window definitions, Chemistry, Chemical Engineering, and Physics and Astronomy, are exactly fields of this second kind; Physics and Astronomy, where research is organized around large instrument-based collaborations, shows no rebound at all.
What matters is the substitutability of the execution work, not team-based organization as such: Computer Science and Psychology are organized around the laboratory model of large, hierarchical, grant-funded teams~\cite{tripodi2025tenure}, yet both reverse strongly.
Arts and Humanities, where solo authorship has remained the norm and thus leaves little room for a substitution-driven rebound, continues to decline. This ordering parallels the documented field gradient in LLM uptake, which is fastest in computer science and adjacent computational fields~\cite{kobak2025, liang2025quantifying}, and it points ahead to the content-level evidence to be discussed below, where recovered solo papers tilt toward computational work.
\subsection*{Author composition alone does not explain the trend break}
\label{sec:within}
\begin{figure*}[t]
  \centering
  \includegraphics[width=\textwidth]{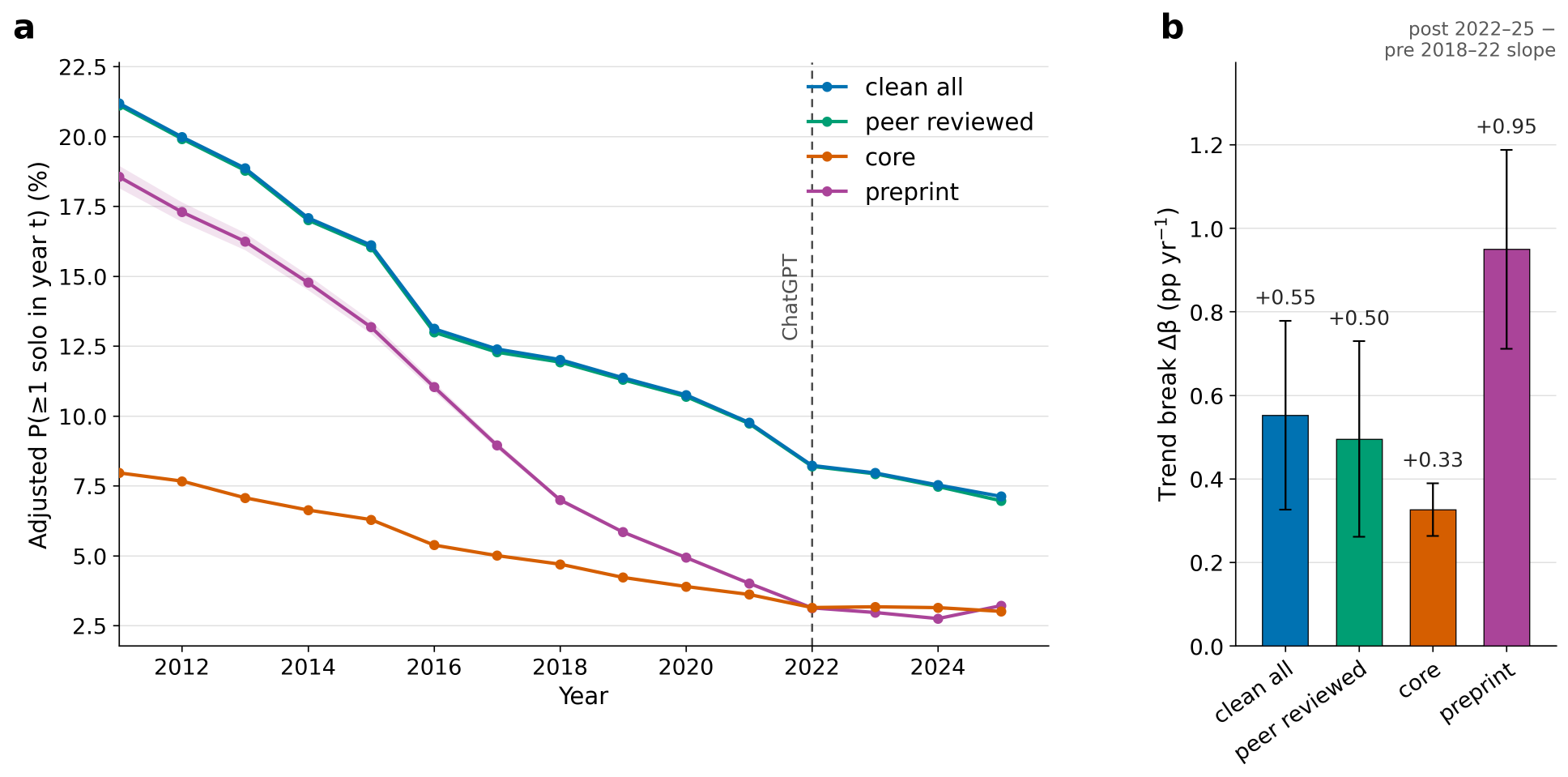}
  \caption{%
    \footnotesize
    Author-level (marginal) probability that an active author publishes at least one solo-author paper in year $t$, and its 2022 trend break.
    \textbf{(A)} Composition-adjusted $P(\geq\!1\text{ solo in year }t)$ by calendar
    year (2011--2025), among authors with at least one publication in year $t$ under
    the given publication-filter variant (clean all, peer reviewed, core, preprint);
    the dashed vertical line marks the public release of ChatGPT (Nov.\ 2022). Each
    curve nets out the changing composition of active authors (field $\times$
    academic age $\times$ lifetime citations $\times$ productivity) via a saturated
    linear probability model, so the plotted movements hold the observable author mix
    fixed.
    \textbf{(B)} Trend break $\Delta\beta$ per variant: the post-2022 minus pre-2022
    weighted-least-squares slope (pre 2018--2022, post 2022--2025; 2022 enters both
    windows) in percentage points per year; whiskers are 95\% confidence intervals.
    }
  \label{fig:within_pillar3-marginal-ab}
\end{figure*}
Yet a rise in the aggregate solo-authorship rate can arise through two distinct channels: compositional change and within-author change. The first channel, compositional change, refers to a turnover in the set of researchers who write papers.
To separate these two channels, building on scholar-level analyses of solo-authorship dynamics~\cite{kwiek2022female,lazebnik2025lonely}, we condition authors on their past solo-authorship history and follow what happens within groups of authors defined by that history. To do so, we define the probability that a given author publishes a solo-authored paper (Materials and Methods). By explicitly varying the threshold $k$, the number of active years without a solo-authored paper, we ensure the robustness of the result. A higher $k$ retains only authors with no solo paper in their recent active years, including those who have never published solo, and it allows us to examine change conditional on individual authors' own histories rather than the aggregate solo-authorship rate.
Carrying out this computation, we find that even after restricting authors by history, such as the length of their purely coauthored record or whether they have never written solo at all, the decline in the probability of solo authorship still halts or reverses (Fig.~2 and SI Fig.~S5). The history-conditioned solo-authorship rate estimated in this way had also declined on average over the past two decades. After the spread of LLMs, however, the trend halts or reverses.
This result is robust in an analysis that adjusts for the changing author mix, netting out field, academic age, lifetime citations, and productivity (Fig.~2). After a steady pre-2022 decline of roughly $-0.9$ percentage points per year ($-0.4$ for core), the composition-adjusted probability halts or reverses at 2022 under every filter variant, with the trend break positive. We obtain the same reversal after holding field, academic age, lifetime citations, and productivity constant, indicating that compositional changes along these dimensions do not explain the result.
We further examine how the reversal appears as we tighten the conditioning. Raising the threshold from $k\!=\!1$ to $k\!=\!15$ lengthens the required run of coauthored-only activity, and the conditioned population becomes increasingly dominated by authors who have never published solo; at $k\!=\!15$ the estimate approximates the hazard of a first-ever solo-authored paper (Materials and Methods). The post-2022 halt or reversal is present at every threshold and, if anything, strengthens with $k$ (SI Fig.~S5). The effect is therefore at least as strong among the authors least accustomed to publishing alone. As noted above, field-level and author-level breaks need not coincide, however: in Engineering, where the share-level break is the largest, the author-level probability barely moves, so the aggregate reversal there appears compositional rather than within-author (SI, Section~H).
\subsection*{The mechanism is consistent with LLM substitution}
If LLMs are substituting for coauthors, the reversal should carry two signatures.
First, who moves: the shift to solo authorship should be strongest among authors for whom LLMs most reduce the cost of publishing without human help, not only among those already close to writing alone.
Second, what they write: authors moving to new topics would indicate a shift in interest~\cite{zeng2019topics}, whereas authors writing alone about content they previously addressed with coauthors would indicate that LLMs are supplying the labor coauthors once provided.
We examine each in turn.
The profile of who moves to solo authorship matches the substitution view (SI Fig.~S6).
When authors are grouped by lifetime productivity, the 2022 rebound appears in every group, but it is largest among the least prolific authors and smallest in the mid-productivity range (6–20 works).
This pronounced peak among low-output authors is consistent with LLMs lowering the fixed cost of producing a paper without a human coauthor.
Among authors with no recent solo publication ($d_{\text{active}}\!\geq\!3$), moreover, the rebound is present in every seniority class, and under the main filters the largest break appears among the most senior researchers, although the age gradient is weak (SI Fig.~S7).
Seniority is therefore not a barrier: authors who have long divided execution work among coauthors also return to solo authorship once that work can be handed to an LLM.
These findings indicate that what has changed is not simply the number of papers, but with whom, or without whom, research is completed.
\subsection*{Going solo shifts content toward computational work}
\label{sec:content-shift}
Having established who moves to solo authorship, we turn to what they write.
The difficulty is that the content of science as a whole has been moving since 2022, so a naive before–after comparison of solo papers would attribute to solo authorship what is in fact a field-wide trend.
We therefore compare four cohorts of papers, solo-authored and team-authored, before (2018–2021) and after 2022, embedded with SPECTER2, and measure the solo-specific displacement as a difference-in-differences (Materials and Methods).
The displacement points in a specific direction rather than a random one.
Net of field-wide drift, post-2022 solo output concentrates in specific regions of the content space and moves out of others (Fig.~3a; 1{,}387 of 5{,}883 populated cells of the two-dimensional atlas significant at the 95\% level).
Projected onto four pre-specified, keyword-anchored semantic axes ($n = 305{,}231$ English-language papers; SI, Section~I), solo papers shift toward the computational end ($+0.040$~s.d., $P = 4\times10^{-8}$), whereas the data-rich–versus–data-scarce ($+0.016$~s.d., $P = 0.15$), review-versus-original ($+0.014$s.d., $P = 0.07$), and theoretical-versus-applied ($+0.010$s.d., $P = 0.19$) axes show no significant movement (Fig.~3b).
Only one of the four axes moves significantly, which shows that the shift is specific: by this measure, the recovered solo papers are ordinary original research that has tilted toward computation.
Going solo therefore does not move researchers to a different region of science. Rather, on top of a drift shared with everyone else, solo output leans measurably further toward computational work.
The content signature of the solo rebound is thus not concentrated in a handful of topics but appears as a broad tilt across many regions of the content space (Fig.~3a) toward the kind of work for which LLM assistance is strongest.
The cohort-level displacement has an author-level counterpart.
We compare what researchers write alone with what they write with coauthors, rather than tracking who abandons collaboration, so an author who publishes both solo and coauthored papers after 2022 enters both groups (SI, Section J).
Comparing authors of post-2022 solo papers (solo-writers) with authors of post-2022 coauthored papers (persistent coauthors), the content breadth of solo-writers’ own post-2022 papers, measured as the mean pairwise cosine distance among an author’s own papers, is 0.089 versus 0.115, that is, 23\% narrower (regression-adjusted difference $-0.028$, $P < 10^{-16}$, $n = 35{,}893$ authors; SI Fig.~S9).
Exploration, by contrast, barely moves.
The distance of solo-writers’ post-2022 papers from their own pre-2022 collaborative centroid is statistically indistinguishable from that of persistent coauthors ($+0.001$, $P = 0.07$, $n = 45{,}642$ authors), an order of magnitude smaller than the breadth effect (SI Fig.~S9b; both models include field fixed effects and paper-count controls).
In other words, solo-writers show no detectable movement toward new territory; they keep working near the content of their coauthored era while narrowing the range of what they attempt alone.
These content patterns discriminate between two readings of the rebound.
If going solo reflected a change of interest, solo-writers should move away from their past coauthored content, and their solo papers need not cluster in any particular region of the content space.
If instead LLMs are taking over execution work that coauthors once supplied, researchers should carry alone a slice of what they previously did with others: staying near their past content, covering less of it, and concentrating on work whose execution layer, such as coding, data handling, and statistical analysis, an LLM can absorb.
The observed pattern, a diffuse tilt toward computational work, no movement toward reviews, and breadth contraction without exploration, matches the second reading on every dimension we measure.
A complementary check using external data supports these findings (SI, Section~K).
\begin{figure*}[!t]
  \centering
  \includegraphics[width=\textwidth]{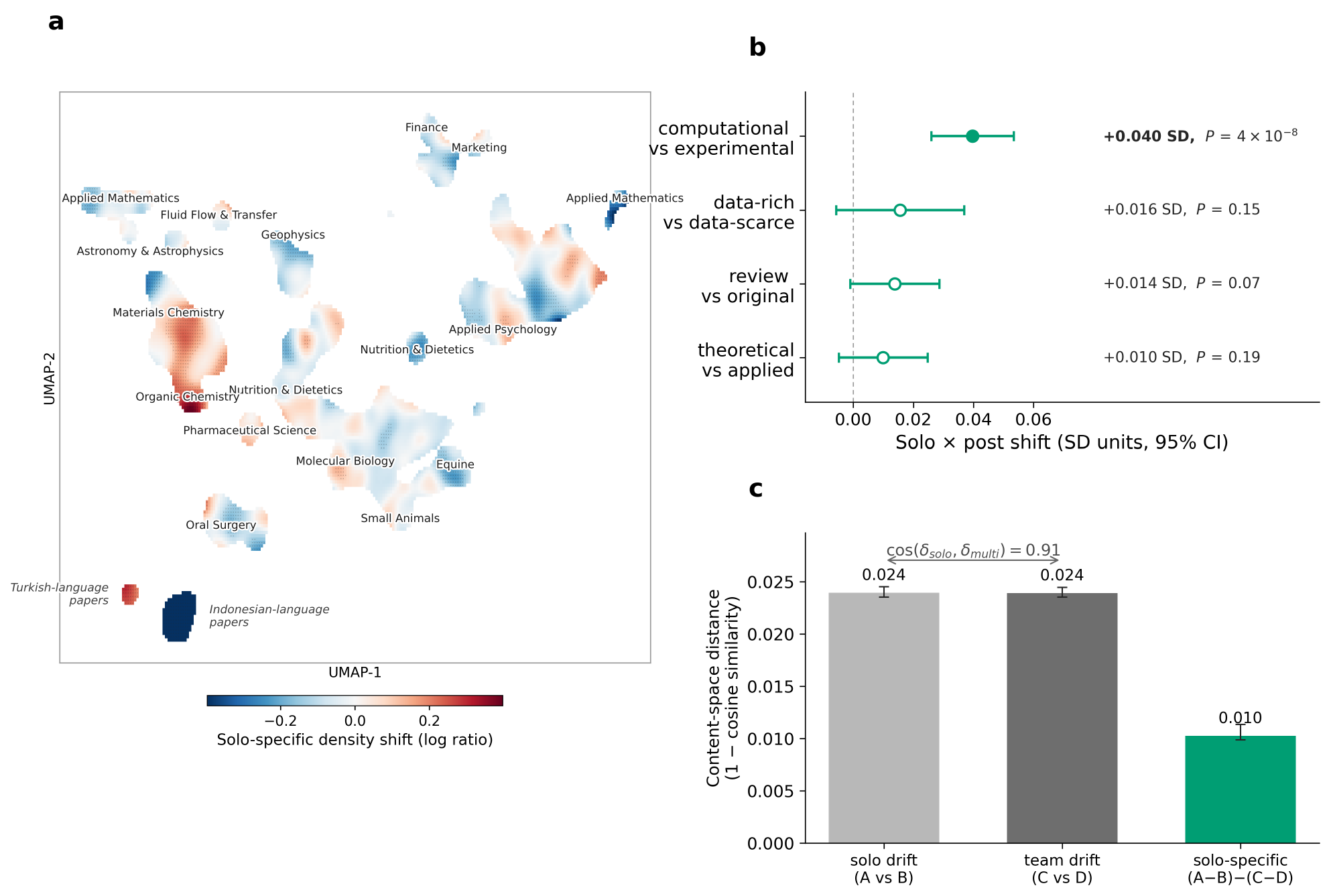}
  \caption{\textbf{Going solo shifts content toward computational
    work: a displacement distinct from field-wide drift.}
    Cohorts: A, solo-authored post-2022; B, solo-authored 2018--2021;
    C, team-authored post-2022; D, team-authored 2018--2021 (cohort
    construction, embedding, and estimation in \textit{Materials and
    Methods}). Displacement statistics in panels b and c are computed in the full
    768-dimensional SPECTER2 embedding space; the two-dimensional atlas
    is a UMAP projection~\cite{McInnes2018} (axis directions arbitrary).
    The density shift and significance cells in panel a are computed on
    this plane and localize where the shift concentrates; the
    quantitative claims rest on the 768-dimensional statistics.
    \textbf{a}, Solo-specific density shift,
    $\log(f_A/f_B) - \log(f_C/f_D)$. Red, regions where solo output
    concentrated after 2022 net of field-wide drift; blue, regions it
    moved out of; stippling, cells whose bootstrap 95\% confidence interval
    (CI) excludes zero. Text labels name the largest regions;
    two labeled regions are language-defined clusters (Turkish- and
    Indonesian-language papers) rather than topics, and they enter the
    map but not the English-only axis analysis in \textbf{b} (SI,
    Section~I).
    \textbf{b}, Difference-in-differences (DiD; solo $\times$ post) of
    each paper's projection onto four keyword-anchored semantic axes,
    in within-cohort s.d.\ units. Dots, point estimates; error bars,
    95\% CI; filled, $P < 0.05$; open, not significant.
    \textbf{c}, Displacement geometry: solo drift (centroid shift,
    A vs.\ B), team drift (C vs.\ D), and the solo-specific residual;
    error bars, bootstrap 95\% CIs (multinomial resampling of each
    cohort's papers, 500 replicates).}
  \label{fig:content-shift}
\end{figure*}
\section*{Discussion}
At first sight, the decades-long rise in the average number of authors and the current recovery of solo authorship may seem incompatible.
The two are not contradictory, however, because the mean and the tail need not move together.
This study has shown the movement that the mean cannot capture: in the left tail, a decades-long decline halts and reverses.
Solo-authored papers remain a small minority of scientific output, and nothing in our results suggests that solo work is displacing teams as the dominant mode of production.
Our claim concerns the direction of change in the left tail, not its level, which continues to sit far below its pre-decline value~\cite{cunningham2025}.
The present study reveals that the decline has halted and the trend has turned, but the lost ground has not been recovered.
In the Introduction we set the acceleration and substitution views against each other.
Our results do not settle this conflict in favor of one view.
Rather, they assign the two views to different parts of the distribution: the substitution view to the left tail, where our evidence is direct, and the acceleration view to the upper tail, which our tail-based design does not measure.
Generative AI does not push science uniformly toward larger teams; it can add opposing forces to the division of cognitive labor, and the force documented here is the one thickening the solo-authored tail.
This is also why the left tail matters beyond its size~\cite{wu2019large}.
Whether and how researchers use LLMs is rarely disclosed and is hard to observe directly.
A solo-authored paper, by contrast, is an observable behavioral trace: it is, by construction, work completed without credited human coauthors, so the set of papers that can be written alone marks the boundary of what can be completed without credited collaborators.
When AI moves that boundary, the movement appears first in the left tail.
Analyzing who moves there and what they write there thus turns solo authorship into an empirical instrument for measuring the substitution of AI for research labor.
\subsection*{Implications}
Two implications follow from these results.
First, the meaning of authorship and credit changes.
A solo-authored paper has so far indicated work that a single researcher completed alone.
If LLMs are taking over the execution work that coauthors once supplied, this interpretation no longer holds.
Previously, the allocation of research tasks within a paper could be observed directly, because contributions were divided among human coauthors and recorded in contribution statements~\cite{Lariviere2016}.
Likewise, sequences of collaboration among humans made it possible to infer how credit is allocated across coauthors~\cite{shen2014}.
For solo-authored papers in which an LLM supplies the execution work, neither approach applies: the contributing agent appears in neither the author list nor the collaboration network.
Contributors who never appear on author lists have long been common, and the gap between authorship norms and the actual division of labor is well documented~\cite{jabbehdari2017}.
LLM contributions widen this gap in a more fundamental way, shifting invisible labor from unlisted humans to machines that cannot be listed at all~\cite{nature2023groundrules}.
As a result, authorship no longer reliably indicates who contributed what.
This substitution may also partially reverse the long-run growth in the size of core research teams~\cite{Milojevic2014}, contracting them toward solo authorship.
Second, the way early-career researchers are trained and build their
careers is affected~\cite{andalon2024teamwork}.
In this study, the return to solo authorship appears among long-established authors as well as newcomers and is, if anything, strongest among the authors least accustomed to publishing alone.
Collaboration, however, has so far been not only a place to produce results but also a place where junior researchers learn on the job, from more senior researchers, how to write and how to analyze.
The more senior researchers can hand execution work to LLMs instead of to junior coauthors, the weaker this training pathway through collaboration may become.
\subsection*{Limitations and future work}
This study has several limitations.
First, we treat the November 2022 release of ChatGPT as a single shock.
We therefore cannot claim that the reversal is the causal effect of LLMs; other changes at the same time, such as the fading of pandemic-era publishing, may contribute.
Our evidence is correlational.
It rests on three patterns that point the same way: field differences in substitutability, conditioning on author histories, and the contrast with non-reversing fields.
We also read the field heterogeneity as an ordering rather than a quantitative test: occupation-based exposure scores~\cite{Eloundou2024, felten2023heterogeneity} do not map cleanly onto the OpenAlex fields, and a cross-section of 26 fields has little statistical power.
The ordering is, however, consistent with independent measures of LLM uptake~\cite{kobak2025, liang2025quantifying} and with the content evidence of Fig.~3, which uses no field labels.
The halt may also be temporary; the decline could resume as journal policies, authorship norms, or the technology evolves.
Even so, the episode keeps its value.
A decline that had persisted for decades paused at the same time across fields, net of the changing author mix, and with a distinctive content signature.
This pattern measures which parts of research can, at this moment, be completed without credited human coauthors.
The contribution is this measurement, not a forecast that solo authorship will keep rising.
Second, this study shows a change in quantity, that solo authorship recovers, and a change in content, that solo papers narrow in scope and shift toward computational work, but it does not address the quality of the recovered solo papers themselves.
Adding an axis that measures this trade-off between quantity and quality is a task for future work.
A related concern is quality and paper-mill inflation: the rebound is largest in the preprint variant, which is consistent not only with substitution but also with an influx of low-cost, possibly LLM-generated solo output.
The rebound, however, is not confined to preprints: it remains positive with 95\% confidence intervals excluding zero under the peer-reviewed and core filters, so an influx confined to preprint venues cannot account for the result; because these filters are venue-based proxies that cannot verify refereeing quality, however, they cannot rule out mill-style output within indexed venues.
At the same time, if part of the preprint excess does reflect mill-style production, this would itself imply that LLMs now enable such output to be produced without coauthors; separating genuine substitution from solo paper-mill inflation within the preprint segment is an interesting direction for future work.
Third, this study rests on a single corpus, OpenAlex, the bulk of which consists of English-language scientific papers.
Whether the change observed here is also seen in other institutional contexts and in non-English-language communities requires separate investigation.
In addition, OpenAlex's own ingestion pipeline and author-disambiguation algorithm have been updated during the observation window~\cite{alperin2024analysis, culbert2025reference}, including the transition away from Microsoft Academic Graph at the end of 2021; the balanced venue panels (SI, Section~L) bound the source-turnover margin of this risk, but metadata changes within continuously indexed venues cannot be fully excluded.

\section*{Materials and Methods}
\subsection*{OpenAlex Data}\label{sec:method_data}
For data we use the full OpenAlex snapshot (1990--2025)~\cite{priem2022openalex}, specifically the release of June 25, 2026.
Because OpenAlex does not carry a complete peer-review flag, peer-review status is operationalized from the work type and the host venue; we run four publication filters in parallel (clean all, peer reviewed, core, preprint), with peer reviewed as the target of the main analysis (precise definitions in SI, Section~A).
Authors are matched to papers via OpenAlex author IDs (algorithmically disambiguated author profiles~\cite{alperin2024analysis}), and each work is assigned to one of the 26 fields of the OpenAlex topic taxonomy through its primary field.
The analysis rests on four metrics: the field-level solo-authorship share (Fig.~1), the mean number of authors per paper (SI Fig.~S2), the author-level solo-authorship probability defined below (Fig.~2), and the content displacement of solo papers (Fig.~3).
\subsection*{Embeddings}\label{sec:method_emb}
For the embedding model, we use SPECTER2~\cite{singh2023spector2}, developed based on SPECTER~\cite{Cohan2020SPECTER}.
We measure the shift of solo output net of the contemporaneous shift of team output (Fig.~3; 636{,}454 papers; the displacement statistics of Fig.~3b,c are computed in the full 768-dimensional embedding space, and the density map of Fig.~3a on its two-dimensional UMAP atlas; SI, Section~I). Under this design, any movement common to solo and team papers is absorbed as field-wide drift, and only the displacement specific to solo output remains.
\subsection*{Author-level solo-authorship probability}\label{sec:method_solo_prob_cond}
To measure solo authorship at the level of individual authors, we constructed an
author--year panel from the OpenAlex snapshot: for each publication-filter variant
(clean all, peer reviewed, core, preprint), an author enters the panel in every
calendar year in which they published at least one paper satisfying that variant,
and the outcome $y_{it}$ indicates whether author $i$ published at least one
solo-authored paper in year $t$; author records with implausible output volume
($>$2{,}000 lifetime works, or $>$50 works per active career year) were excluded.
For each variant we report a composition-standardized annual probability obtained
by direct standardization: author--years were grouped into composition cells
defined by field $\times$ academic-age class $\times$ lifetime-citation class
$\times$ productivity class, the raw probability $p_{ct}$ was computed within each
cell--year, and the annual estimate is the weighted average
$\sum_{c} w_{c}\,p_{ct}$ with fixed weights $w_{c}$ proportional to each cell's
pooled 2010--2025 author-year count, renormalized over cells observed in year $t$.
Field is a time-invariant author attribute: each author is assigned the parent OpenAlex field of the most frequent primary subfield among their pre-2022 publications (ties broken arbitrarily; authors first publishing after 2021 form a residual category). Authors publishing in several fields within a year therefore occupy a single, well-defined cell, and the standardization holds pre-period field affiliation fixed rather than conditioning on post-2022 field choices, which could themselves respond to going solo.
This estimator is numerically identical to the covariate-averaged prediction of a
fully interacted linear probability model with year $\times$ cell indicators, and
therefore holds the observable composition of active authors fixed without
imposing additivity; 95\% bands treat cell counts as independent binomials.
The trend break $\Delta\beta$ is the difference between weighted-least-squares
slopes of the standardized series estimated separately on 2018--2022 and
2022--2025 (weights equal to annual author-year counts; 2022 enters both
windows), expressed in percentage points per year, with standard errors combined
across the two segments.
We define the history-conditioned solo-authorship probability as
\begin{equation}
  P_t(k) = \Pr\!\left(y_{it} = 1 \;\middle|\; d_{it} \ge k\right),
  \label{eq:ptk}
\end{equation}
where $y_{it}$ indicates whether author $i$ publishes a solo-authored paper in
year $t$, and $d_{it}$ is the recency, in active years, of author $i$'s most
recent solo-authored year.
Three ingredients of Eq.\ (1) deserve
precise definition. First, an \emph{active year} is a calendar year in which
author $i$ publishes at least one paper satisfying the publication filter
under consideration; both $y_{it}$ and $d_{it}$ are computed within each
filter variant, and $P_t(k)$ conditions on author $i$ being active in year
$t$. Second, writing $t_1 > t_2 > \cdots$ for author $i$'s active years
strictly before $t$,
\begin{equation}
  d_{it} =
  \begin{cases}
    \min\{\, j : \text{author } i \text{ published solo in } t_j \,\}
      & \text{if such } j \text{ exists},\\[2pt]
    \infty & \text{otherwise},
  \end{cases}
  \label{eq:dit}
\end{equation}
so that $d_{it}=1$ if the immediately preceding active year included a solo
paper, and $d_{it}=2$ if the last solo year lies two active years back. The
counter thus resets after every solo year, and it is predetermined at $t$: it
depends only on the publication history strictly before $t$, never on the
outcome $y_{it}$ itself. Because every active year lying strictly between the
last solo year and $t$ contains coauthored papers only, $d_{it}-1$ equals the
number of consecutive coauthored-only active years preceding $t$; the quantity
labeled $d_{\text{active}}$ in the SI and in the figures is exactly $d_{it}$
as defined here.
Third, conditioning on $d_{it}\ge k$ retains author-years with no solo
publication in the $k-1$ most recent active years, including all authors who
have never published solo; $k=1$ imposes no restriction and reproduces the
marginal series of Fig.~2, while as $k$ grows the conditioned population is increasingly
dominated by never-solo authors, so that at our largest threshold ($k=15$)
$P_t(k)$ approximates the hazard of a first-ever solo-authored paper.
Publication histories from 1990 onward enter the construction of $d_{it}$; the
estimation window is 1995--2025.

\paragraph{Data Availability.}
The OpenAlex dataset is publicly available at \url{https://openalex.org} (snapshot of June 25, 2026). 

\paragraph{Author contributions.}
A.M. conceived the study, designed the research, acquired and analyzed the data, interpreted the findings, and wrote the manuscript.

\paragraph{Competing interests.}
The author declares no competing interests.

\paragraph{Acknowledgments.}
The author is grateful to colleagues and seminar participants for constructive comments and discussions that improved the manuscript. This work was supported by JSPS KAKENHI Grant Number JP22K20159 and JST ERATO Grant Number JPMJER2502. The author also acknowledges OpenAlex for providing the bibliographic data used in this study. The author used Claude Opus 4.8 and Claude Fable 5, developed by Anthropic, to assist with developing and debugging analysis code and revising the manuscript for clarity, structure, and English-language expression. The author reviewed and verified all AI-assisted code and text and takes full responsibility for the final analyses and manuscript.

\clearpage
\begin{center}
  {\Large\textbf{Supporting Information}}
\end{center}

\appendix
\setcounter{figure}{0}
\renewcommand{\thefigure}{S\arabic{figure}}
\setcounter{table}{0}
\renewcommand{\thetable}{S\arabic{table}}

\section{Publication-filter variants}
\label{sec:si_filters}

OpenAlex does not carry a complete peer-review flag, so peer-review status must be operationalized from observable metadata: the work type and the type and identity of the host venue (the source of a work's primary location).
All variants share the same base sample: works published 1990--2025 with an assigned primary field (26 fields).
On this base we define four variants that differ in how strictly they restrict the venue.

\emph{Peer reviewed} (the main specification) requires the work type to be \texttt{article} and the primary host source to be of type \texttt{journal} or \texttt{conference}.
This excludes repositories, books, book chapters, dissertations, datasets, and paratext, but it cannot verify refereeing itself; it is a venue-based proxy.
In addition, we exclude a three-title denylist of bulk-indexed sources whose OpenAlex field assignments are spread far more widely than their editorial scope would suggest: \emph{SHILAP Revista de lepidopterolog\'ia} (a lepidopterology journal that is nonetheless the single largest peer-reviewed source in 2023--24, spread across all 26 fields), the \emph{Scientific Issues of Ternopil National Pedagogical University} (two series of a single university bulletin, likewise spread across all fields), and the \emph{Journal of Bioresource Management}.

\emph{Core} adds to peer reviewed the requirement that the host source belong to OpenAlex's curated core-source list (\texttt{is core}, approximating coverage by curated bibliometric databases such as Scopus and Web of Science).
It is the strictest variant.

\emph{Preprint} is defined as an explicit allowlist of genuine preprint and working-paper servers among repository-type sources: arXiv, bioRxiv, medRxiv, ChemRxiv, Research Square, SSRN, RePEc (including the Munich Personal RePEc Archive and EconStor), the National Bureau of Economic Research, Preprints.org, Authorea, and the discipline-specific OSF servers (PsyArXiv, SocArXiv, EarthArXiv, EngrXiv, and others).
An allowlist is necessary because the raw repository source type is heavily contaminated by bulk-indexed thesis archives, institutional repositories, and data or specimen repositories, which are almost entirely solo-authored and would mechanically inflate the solo share; the generic OSF catch-all and OSF-hosted thesis archives (e.g., Thesis Commons) are excluded on the same grounds.

\emph{Clean all} is the union of peer reviewed (after the denylist) and preprint: the broadest coverage restricted to venues with a recognizable editorial or dissemination function.

Where a figure additionally shows an \emph{all} panel (SI Figs.~S1, S3, and S4), that panel applies no venue restriction at all (every indexed work, including books, datasets, and dissertations) and is displayed only for comparability with prior work based on unrestricted OpenAlex coverage~\cite{cunningham2025}.

\clearpage
\section{Cutoff-based changes in the trend of solo-authored publication rates}
\label{sec:si_cutoff_fields}

For each cutoff year \(c\), we estimate the change in the annual trend of the field-level solo-authored publication share before and after \(c\) (SI Fig.~S1).
The pre-period is defined as \([c-4, \ldots, c-1]\), and the post-period as \([c, \ldots, c+3]\); the cutoff year \(c\) itself thus opens the post-period, so the estimate at \(c=2022\) compares the slope in 2018--2021 with the slope in 2022--2025.
The reported quantity is \(\Delta\beta_c\), the change in slope from the pre-period to the post-period, expressed in percentage points per year.
Positive values indicate that the decline in solo authorship slowed or shifted upward after the cutoff year, whereas negative values indicate that the decline became steeper.
Estimates are obtained from weighted piecewise-linear regressions, using the number of papers in each field-year as weights.
This scan applies a uniform rule in which the cutoff year opens the post-period; because much of 2022 predates the release, estimates that do not depend on this assignment, the monthly split at the release date (Fig.~1) and the donut specifications excluding the transition window (Section~M), serve as checks on this construction.

\begin{figure*}[t]
  \centering
  \includegraphics[width=0.85\linewidth]{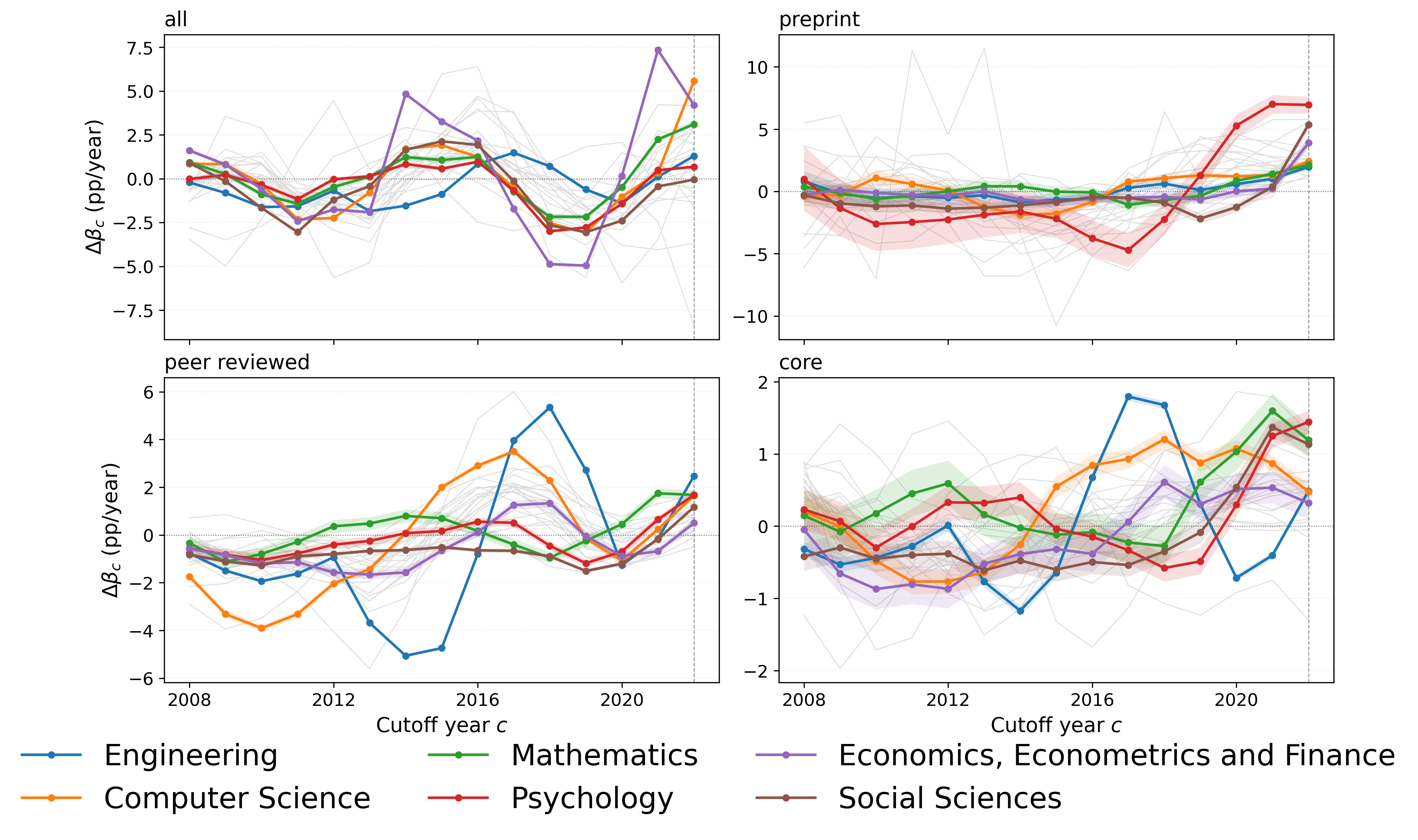}
  \caption{Cutoff-based changes in the trend of solo-authored publication rates.
  Each panel plots \(\Delta\beta_c\), the change in the annual trend of the field-level solo-authored publication share around cutoff year \(c\), estimated with symmetric four-year windows (Section~B).
  Positive \(\Delta\beta_c\) at earlier cutoffs reflect decelerations of a still-ongoing decline, with the post-window slope itself remaining negative; at \(c=2022\), by contrast, the post-window slope turns positive in a broad set of fields.
  Colored lines show the focal fields, while gray lines show other fields for context.
  Shaded bands indicate 95\% confidence intervals based on a parametric binomial bootstrap of solo-author counts within each field-year aggregate; these intervals reflect sampling uncertainty conditional on the observed source universe and field assignments.}
  \label{fig:si_fields}
\end{figure*}

\clearpage
\section{Trend in the mean number of authors per paper}
\label{sec:si_mean_authors}

Separate linear fits of the monthly mean number of authors per paper are estimated before and after November 2022 using weighted least squares with monthly paper counts as weights, after removing calendar-month seasonal effects; 95\% confidence intervals for the fitted trends are based on heteroskedasticity- and autocorrelation-robust (Newey--West) standard errors.
The year 2025 is excluded because of indexing lag.
The reported quantity, $\Delta\beta = \beta_{\mathrm{post}} - \beta_{\mathrm{pre}}$, is the change in the annual slope of the mean author count, expressed in authors per year.
The long-run rise in the mean number of authors flattens or turns downward after the release in most fields and variants ($\Delta\beta < 0$), directionally consistent with the reversal in the solo-authorship share (Fig.~1); the break in the mean is, however, less uniform, with the preprint variant showing essentially no slope change in several fields (SI Fig.~S2).

\begin{figure*}[t!]
\centering
\includegraphics[width=0.8\textwidth]{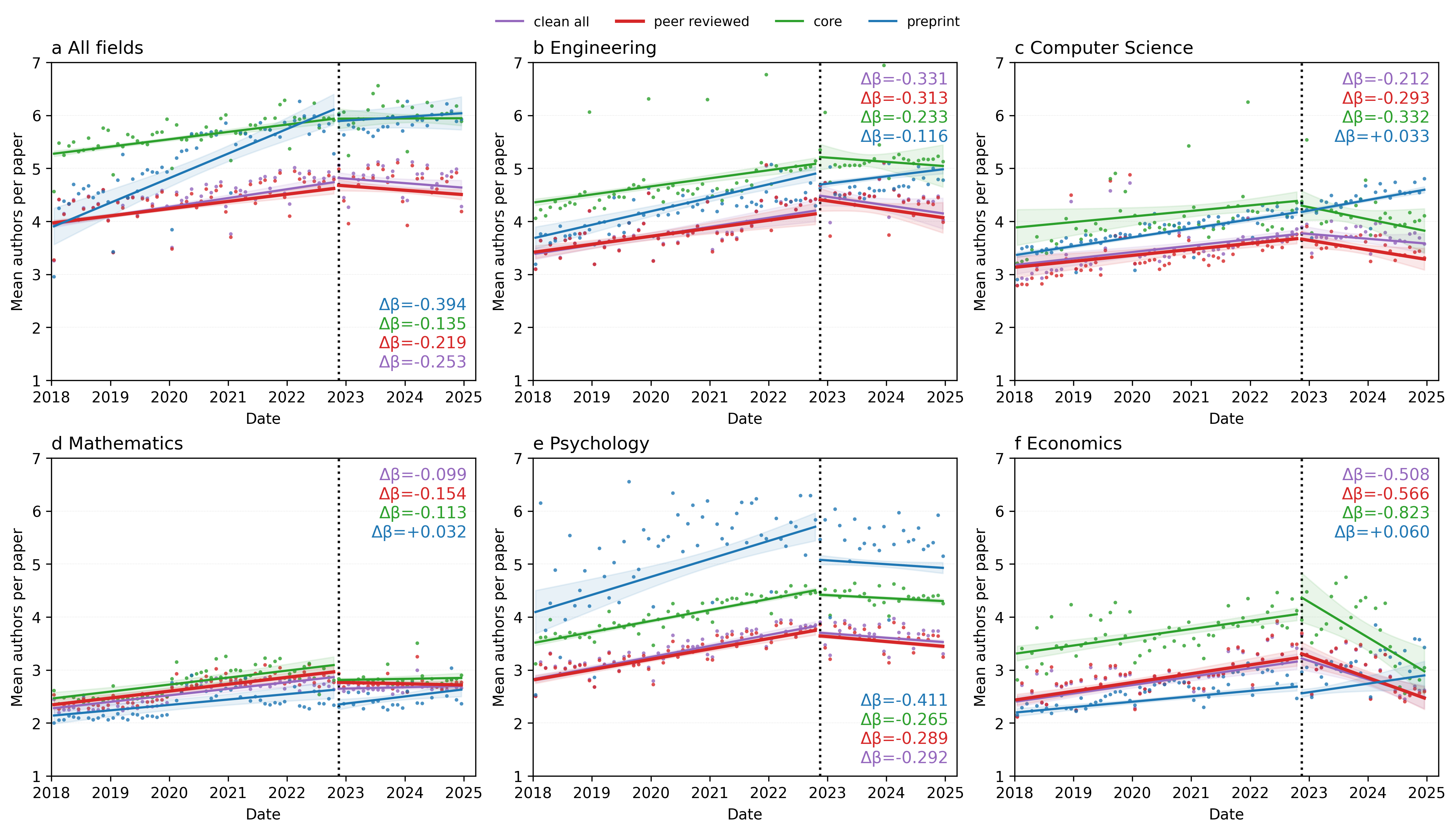}
    \caption{Piecewise-linear trend in the monthly mean number of authors
    per paper at the public release of ChatGPT.
    Panels show the monthly mean number of authors per paper from January 2018 through December 2024 for all fields pooled and for five focal fields: Engineering, Computer Science, Mathematics, Psychology, and Economics.
    Colors distinguish four coverage
    variants, with \emph{peer reviewed} used as the main specification and the
    other variants shown as robustness checks. Dots are monthly observations; the
    vertical dotted line marks November 2022, when ChatGPT was publicly released.
    Estimation details are given in Section~C.}
\label{fig:mean_authors_break}
\end{figure*}

\clearpage
\section{Cutoff-based heatmap of changes in the trend of solo-authored publication rates and the pandemic}
\label{sec:si_si_rolling_symk3}

Each cell of SI Fig.~S3 reports $\Delta\beta_c$, computed with the symmetric windows of Section~B and reported in percentage points per year; cutoff years 2018--2022 are displayed, and 2026 is excluded as an incomplete year.

The heatmap speaks against a pandemic-rebound account, in which the recovery would merely reflect solo shares reverting to trend after the expansion of large-team, COVID-related publishing.
A pandemic dip is indeed visible: at $c=2020$, 20 of the 26 fields show negative slope changes under the main peer-reviewed specification.
Because the post window at $c=2020$ spans 2020--2023 and thus includes the 2022 transition, and because $\Delta\beta_c$ is a slope change rather than a level dip, the comparison below is suggestive rather than a clean measure of pandemic-induced decline; the composition of the recovery, examined in Sections~F and~G, does not depend on this window.
Reversion to trend predicts that the fields that dipped most should rebound most, led by the biomedical fields in which pandemic collaborations were concentrated.
Neither holds.
Arts and Humanities and Chemistry dip deeply ($-1.3$ and $-1.2$~pp/yr) yet do not reverse in 2022, while Mathematics and Decision Sciences show no dip at all ($+0.5$ and $+0.8$) yet sit in the top cluster of 2022 breaks; where a deep dip does precede a large break, as in Engineering ($-1.3$, then $+2.5$), the break is roughly twice the dip, overshooting mere reversion to the pre-pandemic trend.
Nor is the recovery biomedical-led: the largest 2022 breaks occur in Engineering and Business, followed by Mathematics, Decision Sciences, Psychology, and Computer Science, rather than in Medicine or Immunology.
\begin{figure*}[t]
  \centering
  \includegraphics[width=0.85\linewidth]{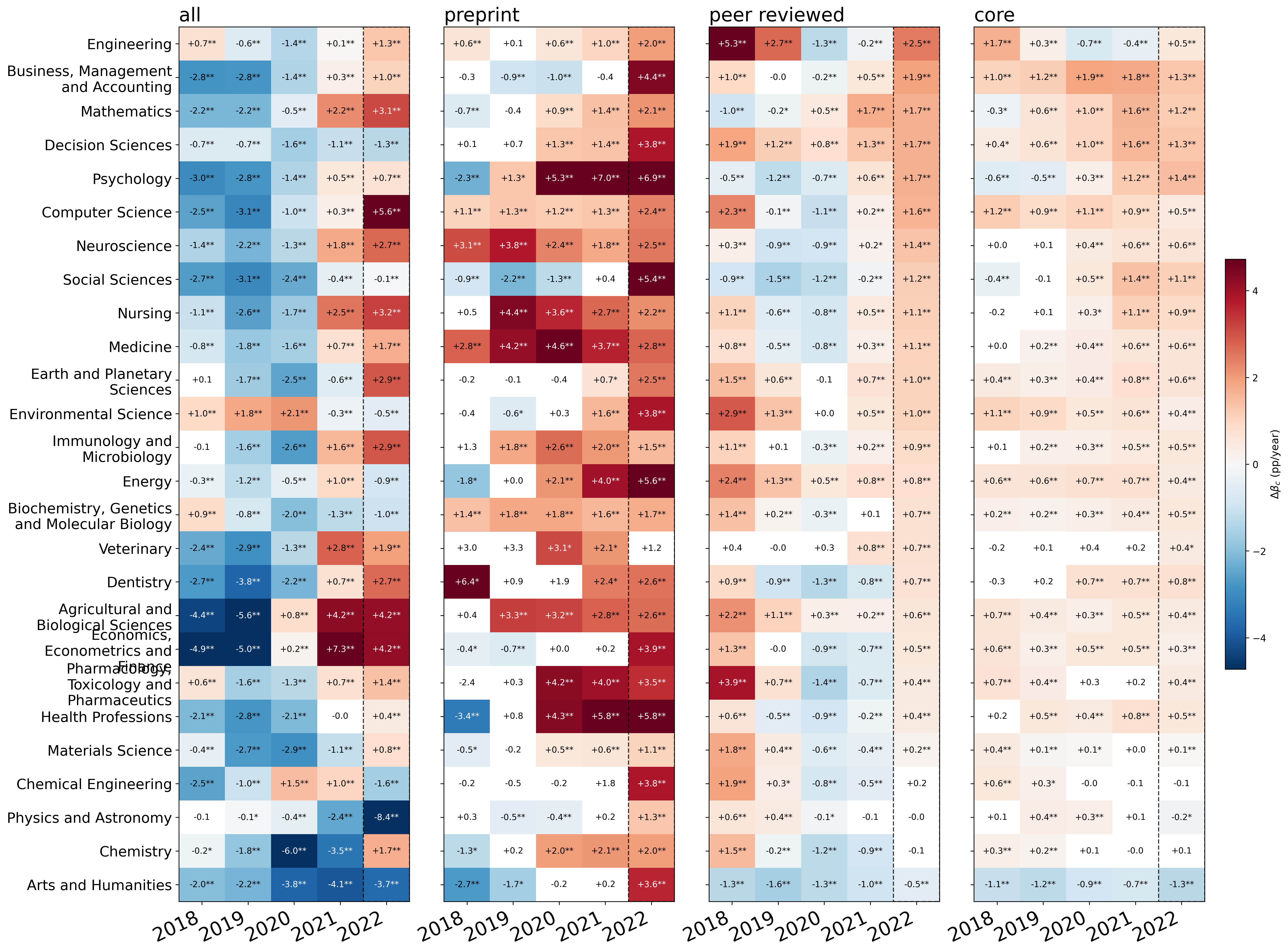}
  \caption{
    Cutoff-based heatmap of $\Delta\beta_c$, the change in the annual trend of the field-level solo-authored publication share around cutoff year $c$ (Section~D).
    Rows represent fields, columns represent cutoff years, and panels represent filter variants, corresponding to all papers, preprints, peer-reviewed papers, and core papers.
    Fields are ordered by their estimated $\Delta\beta_c$ in the peer-reviewed variant at the primary cutoff year $c=2022$.
    Cells are annotated with point estimates, and asterisks indicate bootstrap significance levels, with * denoting $p<0.05$ and ** denoting $p<0.01$; cells that are not significant at $p<0.05$ are shown with a white background.
    These intervals reflect sampling uncertainty conditional on the observed source universe and field assignments.
    The dashed box highlights the 2022 cutoff column.
  }
  \label{fig:si_rolling_symk3}
\end{figure*}

\clearpage
\section{Robustness of the field ordering to the window definition}
\label{sec:si_asym_robust}

This ordering is broadly robust to the window definition.
Under the asymmetric window, which compares the slope in $[\tau\!-\!1,\tau]$ with the long-run slope in $[\tau\!-\!7,\tau\!-\!1]$, largely the same fields lead at $\tau=2023$, the first window ending in a fully post-release year (its recent slope, over $[2022,2023]$, includes the 2022 transition year): Engineering ($+4.3$), Computer Science ($+3.3$), and Psychology ($+2.7$), whereas Chemistry and Chemical Engineering remain at zero ($-0.1$, $+0.1$) (SI Fig.~S4).
Unlike the symmetric cutoff definition in SI Fig.~S3, the post-period here is short and the baseline is long, so the two definitions share no window construction.

\begin{figure*}[t]
  \centering
  \includegraphics[width=0.85\linewidth]{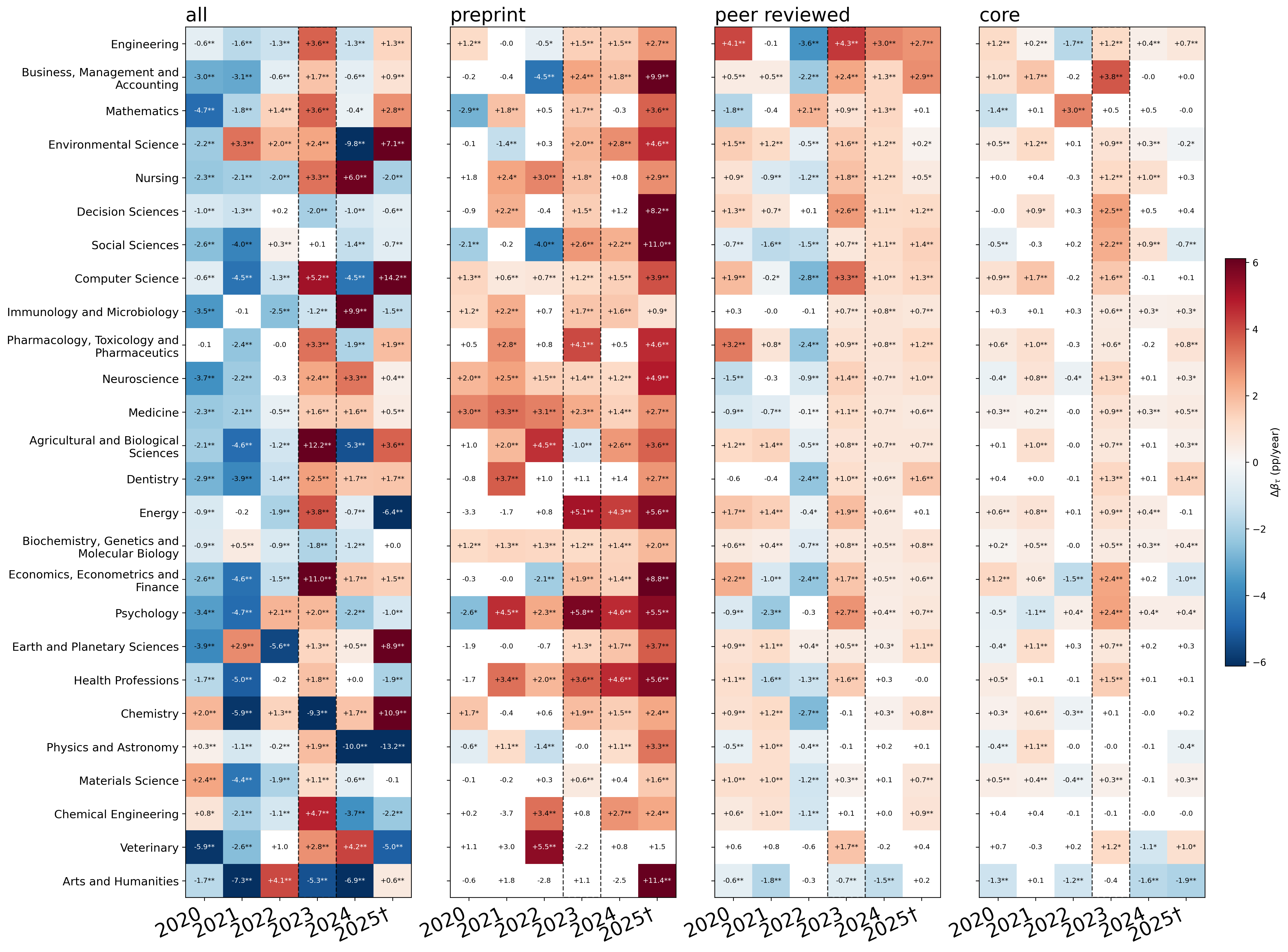}
    \caption{Rolling trend-break $\Delta\beta_\tau$ under the asymmetric window
        definition, as a robustness check on the window choice.
        Each cell reports
        $\Delta\beta_\tau=\mathrm{slope}[\tau\!-\!1,\tau]-\mathrm{slope}[\tau\!-\!7,\tau\!-\!1]$,
        the difference between the slope of the field-level solo-authored
        publication share over the most recent two years and its slope over the
        preceding seven-year baseline, expressed in percentage points per year.
        Positive values indicate that the recent trend turned upward relative to
        the long-run decline.
        Rows represent fields, columns represent window end years
        $\tau=2020$--$2025$, and panels represent filter variants (all, preprint,
        peer reviewed, core).
        Cells are annotated with point estimates; asterisks indicate bootstrap
        significance levels, with * denoting $p<0.05$ and ** denoting $p<0.01$,
        and cells not significant at $p<0.05$ are shown with a white background.
        The dashed box highlights $\tau=2023$, the first window ending in a
        fully post-release year (its recent slope, over $[2022,2023]$, includes
        the 2022 transition year); under the
        main peer reviewed variant, this is the column in which the estimates
        turn broadly positive across fields.
        Estimates at $\tau=2025$ (marked \dag) are sensitive to indexing lags in the most recent OpenAlex records and should be read with caution.
        }
  \label{fig:si_rolling_asym}
\end{figure*}

\clearpage
\section{Solo-authorship probability by switch-in threshold}
\label{sec:si_switchin}

To show that the rebound is not confined to authors already close to solo publishing, authors are restricted by the recency of their last solo-authored year, $d_{\text{active}}\geq k$ for $k\in\{1,2,3,5,10,15\}$: the condition retains author-years with no solo publication in the $k-1$ most recent active years, including authors who have never published solo (Materials and Methods); higher $k$ lengthens the required coauthored-only run, and the conditioned population is increasingly dominated by never-solo authors, so that $k\sim15$ approximates the hazard of a first-ever solo-authored paper.
Each conditioned series is standardized over the same composition cells (field $\times$ academic age $\times$ lifetime citations $\times$ productivity) as the marginal series, so the $d_{\text{active}}\geq1$ series is identical to the marginal series (Fig.~2) by construction.
The post-2022 halt or reversal is present at every threshold and, under the main filters, strengthens with $k$: $\Delta\beta$ rises from $+0.55$ ($d_{\text{active}}\geq1$) to $+0.84$ ($\geq15$) for clean all and from $+0.50$ to $+0.79$ for peer reviewed, while it stays roughly flat under core ($+0.33$ to $+0.26$) and preprint ($+0.95$ to $+0.71$) (SI Fig.~S5).
Because the effect is at least as strong among authors with no recent, or no prior, solo publication as among the rest, the solo rebound reflects authors newly switching into solo work rather than increased output by habitual solo authors.

\begin{figure*}[t]
  \centering
  \includegraphics[width=0.85\linewidth]{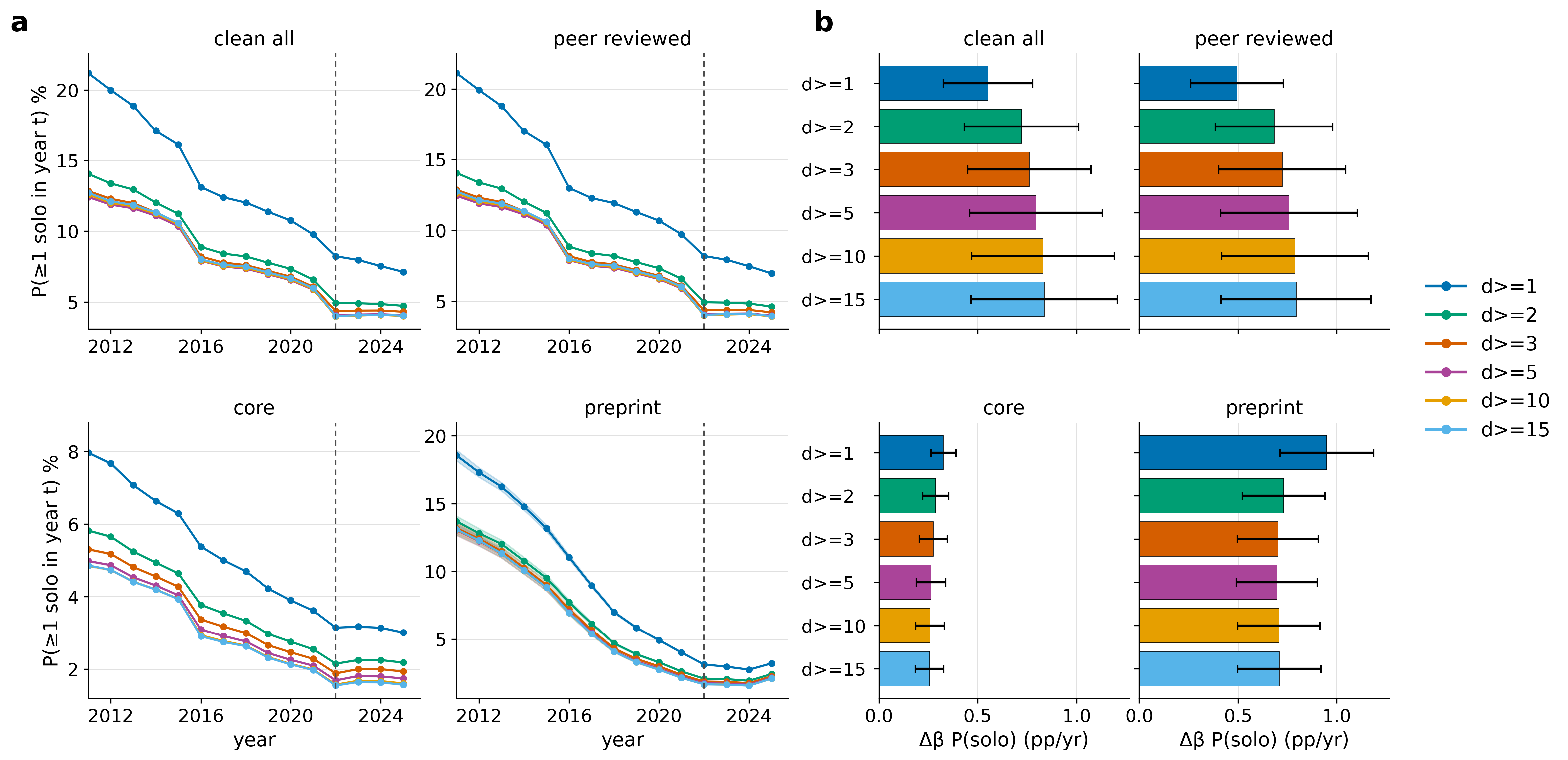}
  \caption{%
    Solo-authorship probability by switch-in threshold $d_{\text{active}}\geq k$
    (Section~F).
    \textbf{(A)} Composition-adjusted $P(\geq\!1\text{ solo in year }t)$ by calendar year,
    a $2\times2$ grid of panels per filter variant, colored by threshold; the dashed line
    marks ChatGPT (2022).
    \textbf{(B)} Trend break $\Delta\beta$ (post-2022 minus pre-2022 WLS slope, pp/yr; 95\%
    CIs) for each threshold.}
  \label{fig:pillar3-switchin}
\end{figure*}

\clearpage
\section{Solo-authorship probability by researchers' productivity and academic age}
\label{sec:si_productivity_age}

When authors are binned by lifetime productivity class (1--5, 6--20, 21--100, and 100+ works), the 2022 rebound appears in every class but is far from uniform: for clean all it is largest among the least prolific authors (1--5 works, $\Delta\beta=+1.16$), smallest for the 6--20 group ($+0.21$), and intermediate for the most prolific (100+ works, $+0.48$) (SI Fig.~S6).
The pattern is most extreme for preprints, where low-output authors show $\Delta\beta=+2.83$~pp/yr.
In the core papers, we also find that the long decline in solo authorship among the most productive authors flattens abruptly after 2022.
The strong low-productivity signal is consistent with LLMs lowering the fixed cost of producing a paper without a (human) coauthor.

Restricting to authors with no solo publication in their two most recent active years ($d_{\text{active}}\geq3$) and splitting by academic age (years since first publication: 1--2, 3--5, 6--10, 11--15, 16--25, 26+), the rebound is present in every age class: for clean all, $\Delta\beta$ ranges from $+0.28$ (6--10 years) to $+0.36$ (26+ years) (SI Fig.~S7).
The oldest class shows the largest break under the clean-all, peer-reviewed, and core filters ($+0.36$, $+0.34$, $+0.24$), but the age gradient is weak and not monotone, and the preprint variant reverses it (largest among the youngest classes, $+0.61$ at 3--5 years versus $+0.40$ at 26+).
Senior scientists, who already have large collaborative networks, thus also return to solo authorship.

Productivity and seniority are different axes, and the seniority analysis conditions on authors with no recent solo publication ($d_{\text{active}}\geq 3$) whereas the productivity analysis does not, so a low-output author need not be junior, nor a senior author prolific.
The two patterns are therefore consistent with each other.

\begin{figure*}[t]
  \centering
  \includegraphics[width=0.85\linewidth]{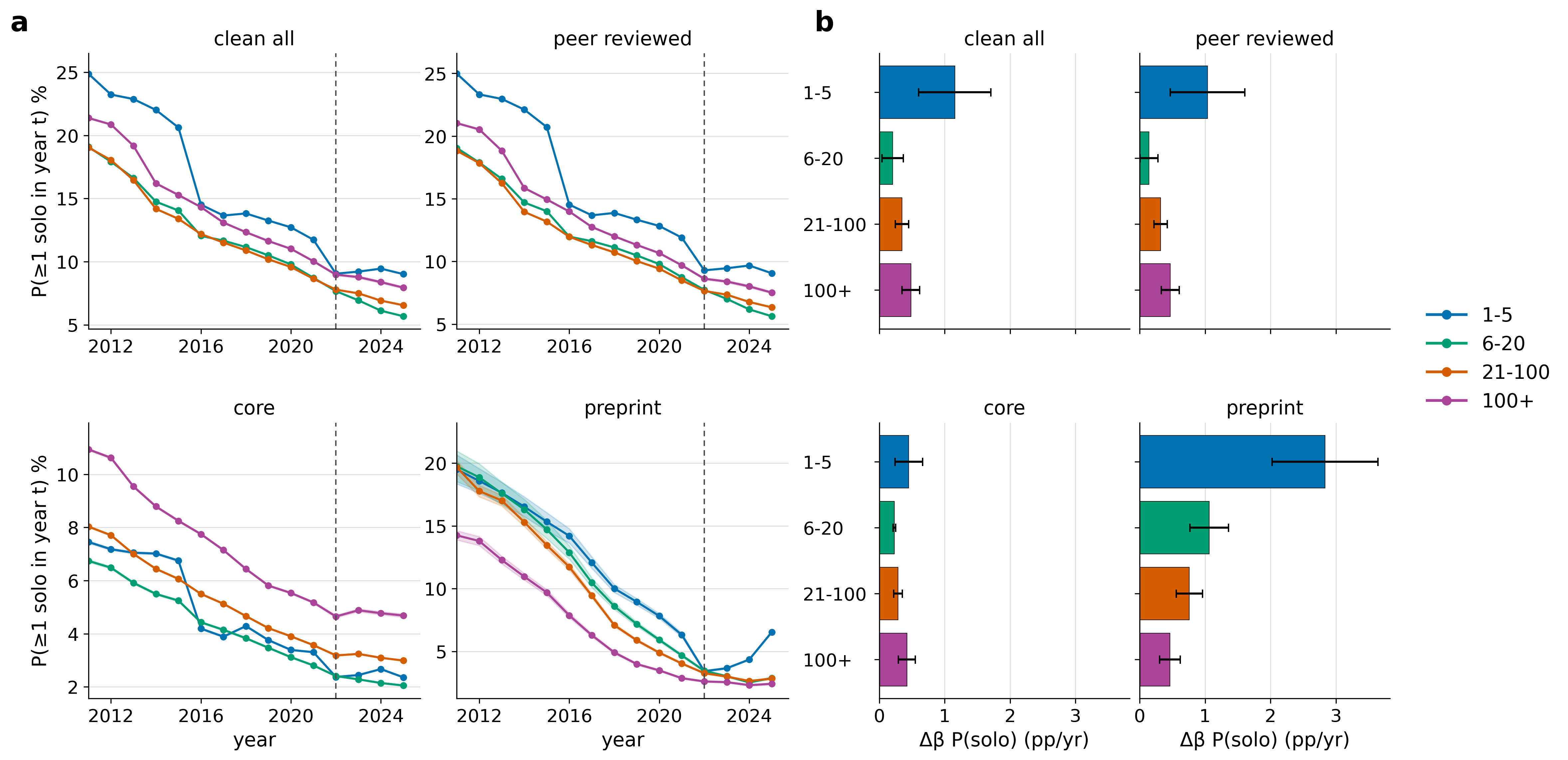}
  \caption{%
    Solo-authorship probability by lifetime productivity class (Section~G). Authors are binned by total
    career output (1--5, 6--20, 21--100, and 100+ works).
    \textbf{(A)} Composition-adjusted $P(\geq\!1\text{ solo in year }t)$ by calendar year,
    $2\times2$ by filter variant, colored by productivity class.
    \textbf{(B)} Trend break $\Delta\beta$ (post-2022 minus pre-2022 WLS slope, pp/yr; 95\%
    CIs) per class.}
  \label{fig:pillar3-nworks}
\end{figure*}

\begin{figure*}[t]
  \centering
  \includegraphics[width=0.85\linewidth]{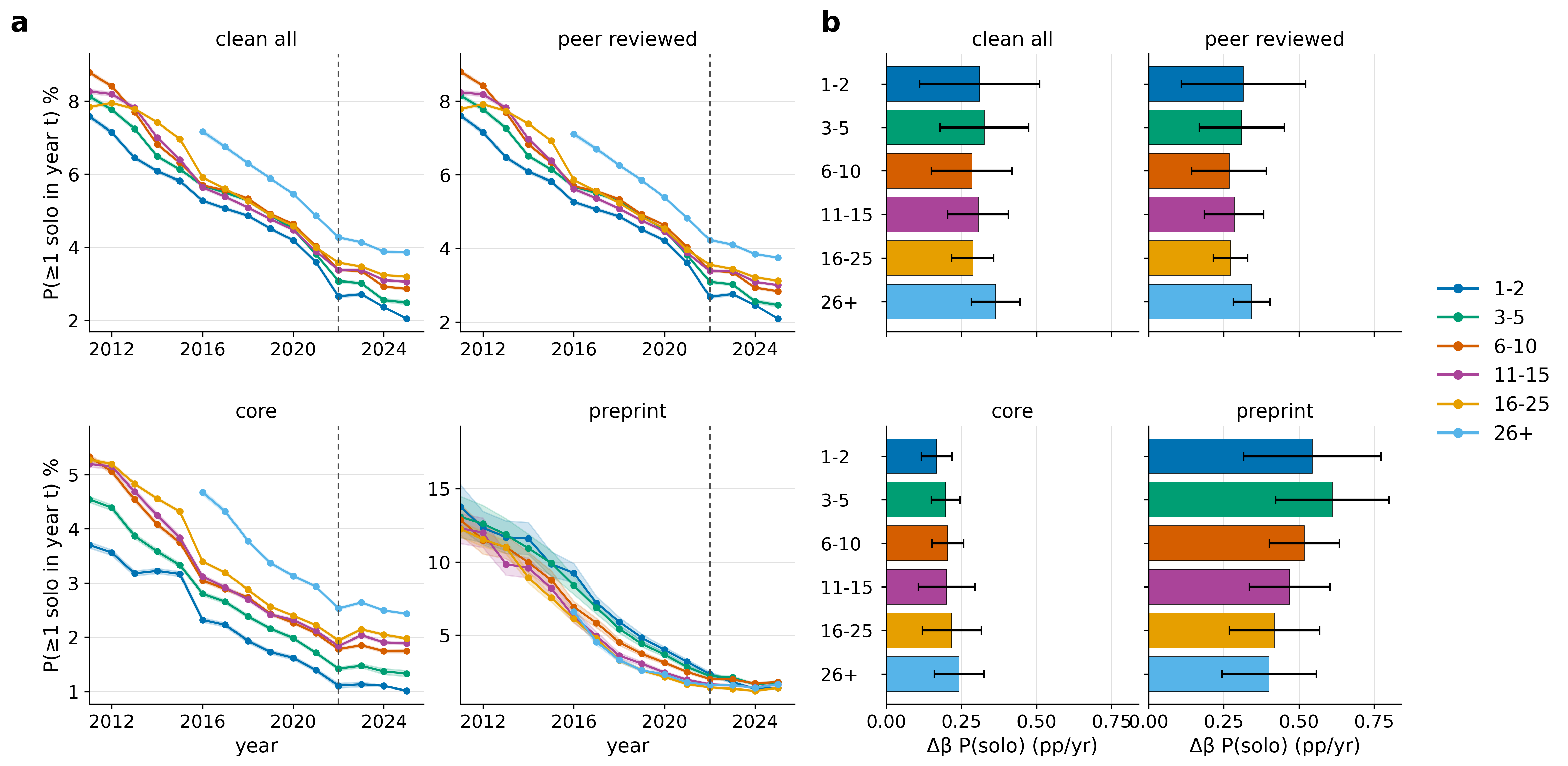}
  \caption{%
    Solo-authorship probability by academic age among authors with no recent solo
    publication ($d_{\text{active}}\geq3$; Section~G). Authors are split by academic
    age (years since first publication: 1--2, 3--5, 6--10,
    11--15, 16--25, 26+).
    \textbf{(A)} Composition-adjusted $P(\geq\!1\text{ solo in year }t)$ by calendar year,
    $2\times2$ by filter variant, colored by age class.
    \textbf{(B)} Trend break $\Delta\beta$ (pp/yr; 95\% CIs) per age class.}
  \label{fig:pillar3-age}
\end{figure*}

\clearpage
\section{Author-level 2022 trend break by field and publication-filter variant}
\label{sec:si_field_dbeta}

The author-level 2022 trend break in the solo-authorship probability is positive in almost every field\,$\times$\,variant cell and is strongest in Economics ($+0.85$/$+1.27$~pp/yr for clean all/preprint), Psychology ($+0.66$ to $+1.34$), and Computer Science ($+0.75$/$+0.80$ for clean all/peer reviewed) (SI Fig.~S8).
The only non-positive cells are Engineering under the clean-all and peer-reviewed filters ($-0.07$, $-0.14$) and Mathematics under the core and preprint filters ($-0.07$, $-0.00$), each with a confidence interval spanning zero.
Mathematics shows a weaker and variant-dependent author-level break despite its strong field-level reversal (SI Fig.~S3), consistent with its long-standing solo norm~\cite{wang2021sciscibook} leaving less room for within-author switching.
The author-level break in Engineering contrasts with its field-level share break, which is the largest across fields (SI Fig.~S3): because the probability that a given Engineering author goes solo barely moves, the aggregate reversal in that field must operate through composition, for example through shifts in the mix of authors or subfields publishing there, rather than through within-author switching.

Elsewhere, this heterogeneity matches the hypothesis that the solo rebound concentrates in fields where LLMs can substitute for a coauthor's language-editing, coding, or statistical role.

\begin{figure*}[t]
  \centering
  \includegraphics[width=0.85\linewidth]{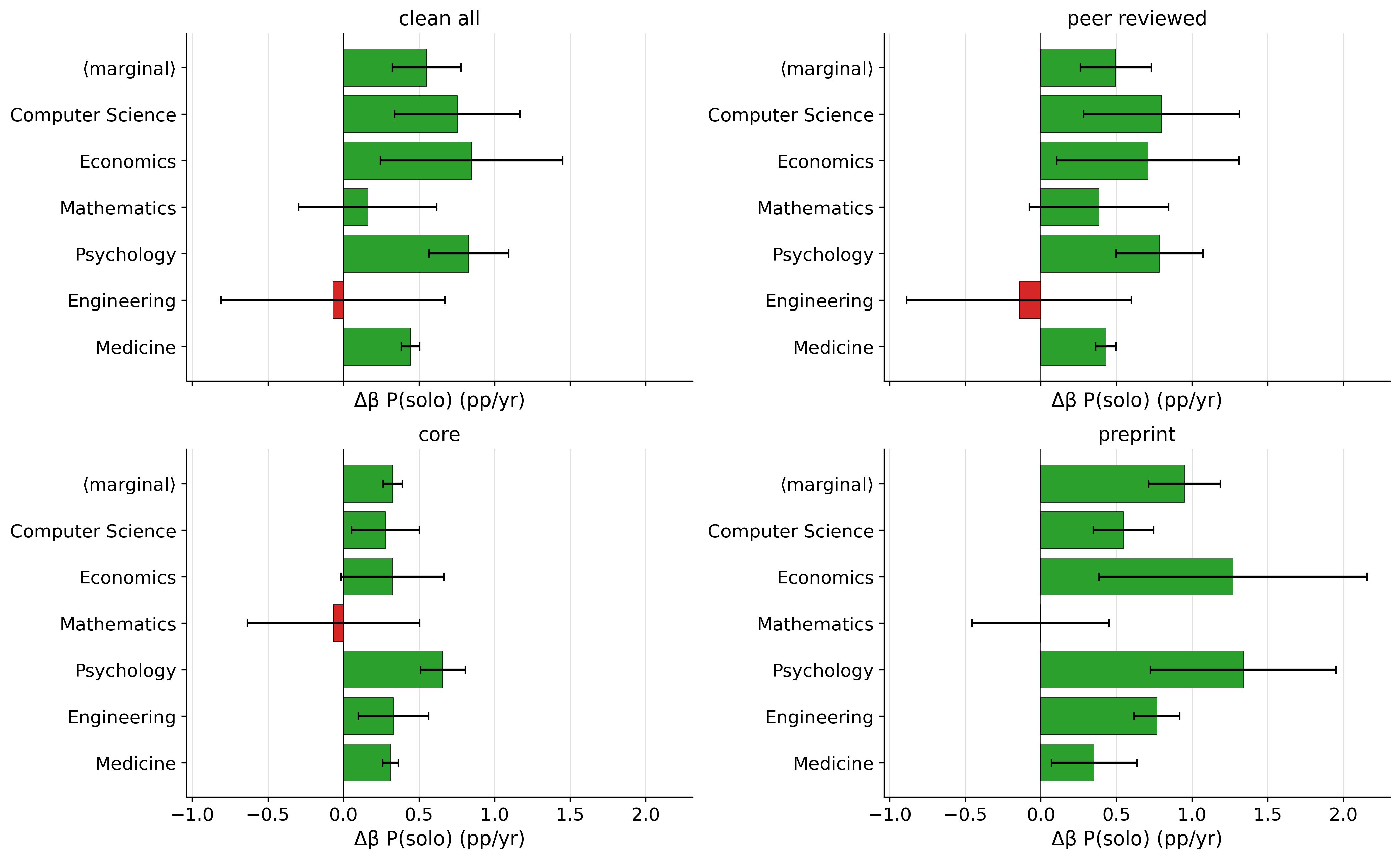}
  \caption{%
    Author-level 2022 trend break in the solo-authorship probability, $\Delta\beta$, by field and publication-filter variant (Section~H). Each
    $2\times2$ panel is a filter variant; within it, horizontal bars give the
    composition-adjusted $\Delta\beta$ (post-2022 minus pre-2022 WLS slope, pp/yr; 95\% CI
    whiskers) for the pooled population ($\langle\text{marginal}\rangle$, i.e.\ not
    conditioned on field) and for six focal fields. Bars are green where $\Delta\beta>0$
    (the pre-2022 decline in solo authorship has halted or reversed) and red where
    $\Delta\beta<0$ (still declining).}
  \label{fig:pillar3-field}
\end{figure*}

\clearpage

\section{Content-shift cohorts, embeddings, and estimation}
\label{sec:method_content}

To measure the content displacement specific to solo authorship (Fig.~3), we constructed four cohorts of peer-reviewed papers: solo-authored post-2022 (A), solo-authored 2018--2021 (B), team-authored post-2022 (C), and team-authored 2018--2021 (D).
Each cohort was drawn by stratified sampling, with up to 5{,}000 papers per field across the 26 fields plus up to 5{,}000 papers for each of the top-20 subfields (strata smaller than the cap enter in full), yielding 636{,}454 papers in total.

Papers were embedded with SPECTER2~\cite{singh2023spector2}, developed based on SPECTER~\cite{Cohan2020SPECTER}.
The displacement statistics of Fig.~3b,c and the author-level analysis of SI Fig.~S9 are computed in the full 768-dimensional embedding space; the density map of Fig.~3a is the only quantity computed on the two-dimensional UMAP~\cite{McInnes2018} plane.

For the density map (Fig.~3a), all embedded papers were binned on a $300 \times 300$ grid spanning the UMAP plane (5\% padding per side), each cohort's counts were smoothed with a Gaussian kernel ($\sigma = 3$ grid cells), the solo-specific shift $\log(f_A/f_B) - \log(f_C/f_D)$ was computed per cell, and cells with a smoothed pooled count below 20 were masked; bootstrap 95\% confidence intervals, obtained by resampling each cohort's papers (multinomial, 200 replicates), identify the significant cells (1{,}387 of 5{,}883 populated cells); at the nominal 5\% level, roughly 294 of the 5{,}883 populated cells would be flagged by chance alone, although kernel smoothing induces spatial correlation across neighboring cells.

In both the density map and the displacement geometry, each cohort's papers are reweighted so that every cohort matches the pooled field composition (direct standardization); movements attributable to a changing field mix are therefore removed before the difference-in-differences is taken.
Because UMAP preserves local neighborhoods rather than global distances or densities, we use the map descriptively, to localize where the solo-specific shift concentrates; no quantitative claim rests on the two-dimensional inference.

For the semantic axes (Fig.~3b), each paper's embedding was projected onto four pre-specified, keyword-anchored axes; the keyword sets, the axis construction, and validation checks are reported at the end of this section (Table~\ref{tab:si_axis_keywords}).
The DiD estimate is the solo $\times$ post interaction in within-cohort s.d.\ units, with field $\times$ year fixed effects and standard errors clustered by subfield ($n = 305{,}231$ English-language papers from the 26-field strata; two-sided tests, $P$ values uncorrected across the four pre-specified axes).
The density map includes papers of all languages and both sampling strata, whereas the axis analysis is restricted to English-language papers of the main field strata; the language-specific regions visible in Fig.~3a (e.g., Turkish- and Indonesian-language papers) therefore enter the map but not the axis estimates.

For the displacement geometry (Fig.~3c), solo drift $\lVert\delta_{\mathrm{solo}}\rVert$ is the centroid shift of A versus B and team drift $\lVert\delta_{\mathrm{multi}}\rVert$ that of C versus D, with distance defined as $1 - \text{cosine similarity}$; confidence intervals are bootstrap 95\% CIs obtained by multinomial resampling of each cohort's papers (500 replicates).

\paragraph{Semantic axes: keyword sets and construction.}
Each axis is anchored by two disjoint sets of English-language papers from the pooled cohorts: a paper anchors a pole if its abstract contains at least one of that pole's keywords (case-insensitive literal substring match; Table~\ref{tab:si_axis_keywords}) and none of the opposite pole's.
The axis vector is the difference between the two anchor-set centroids in the 768-dimensional embedding space, normalized to unit length, $v = (\bar{e}_{+} - \bar{e}_{-}) / \lVert \bar{e}_{+} - \bar{e}_{-}\rVert$; every paper $i$ is scored by the projection $s_i = e_i^{\top} v$, and scores are z-standardized on the pooled English-language sample.
Anchor sets are large for all four axes, ranging from $6{,}058$ vs.\ $39{,}989$ papers (theoretical vs.\ applied poles) to $30{,}230$ vs.\ $23{,}235$ (computational vs.\ equipment poles).

\paragraph{Semantic axes: validation.}
Three checks indicate that the axes measure the intended concepts.

First, anchor separation: the two anchor sets of each axis are separated by $1.10$--$2.25$ pooled s.d.\ along their axis ($1.83$ for computational vs.\ equipment); because the anchors define the axis, this check is partly mechanical, and the following two use no keyword information.

Second, field-level construct validity: mean projections by OpenAlex field rank fields exactly as the axis labels predict: on the computational-versus-equipment axis, Computer Science ($+1.14$~s.d.), Decision Sciences ($+1.01$), and Mathematics ($+0.90$) score highest, while Chemistry ($-1.19$), Pharmacology, Toxicology and Pharmaceutics ($-1.15$), and Immunology and Microbiology ($-0.85$) score lowest.

Third, face validity: the highest-projecting papers that match no keyword of either pole are unambiguous instances of the intended concepts: on the computational pole, the top non-anchor papers concern graph-network traffic prediction, federated-learning optimization, and autoencoder-based filtering; on the equipment pole, bioorthogonal catalysts, core-crosslinked nanogels, and composite-resin cytotoxicity assays.

\begin{table*}[t]
\centering
\caption{Keyword sets anchoring the four semantic axes of Fig.~3b.
Matching is case-insensitive literal substring on the English abstract
text; a paper anchors a pole if it matches at least one keyword of that
pole and none of the opposite pole (``ethnograph'' is a stem matching
ethnography/ethnographic).}
\label{tab:si_axis_keywords}
\small
\begin{tabular}{p{0.17\linewidth}p{0.38\linewidth}p{0.37\linewidth}}
\hline
Axis & Positive pole & Negative pole \\
\hline
review / synthesis $\leftrightarrow$ original empirical &
systematic review; meta-analysis; literature review; scoping review;
narrative review; this review; we review; state of the art; survey of &
we conducted; we performed; randomized; we collected; field experiment;
laboratory experiment; we measured; we estimated; our experiments \\
\addlinespace
computational / desk $\leftrightarrow$ experimental / equipment &
simulation; algorithm; neural network; deep learning; computational;
machine learning; numerical model; software; solver &
apparatus; specimen; in situ; microscopy; spectrometer; reactor; assay;
fabricated; synthesized; clinical trial; field measurements \\
\addlinespace
theoretical $\leftrightarrow$ applied &
theorem; proof; lemma; corollary; axiom; we prove; theoretical framework;
conceptual model &
case study; implementation; practical; application of; deployment;
industry; policy implications; in practice \\
\addlinespace
data-rich $\leftrightarrow$ data-scarce / qualitative &
dataset; big data; database; administrative records; large-scale data;
panel data; corpus of; benchmark &
interview; ethnograph; focus group; small sample; case report; archival
sources; fieldwork; participant observation \\
\hline
\end{tabular}
\end{table*}

\clearpage
\section{Author-level content scope: solo-writers versus persistent coauthors}
\label{sec:si_author_focus}

At the author level, we compare two groups of researchers defined by their
observed authorship in the embedded cohort samples, using each author's own
post-2022 papers in the 768-dimensional SPECTER2 space (SI Fig.~S9).
\emph{Solo-writers} are the authors of the sampled post-2022 (2022--2025)
solo-author peer-reviewed papers (cohort A).
\emph{Persistent coauthors} are all listed authors of the sampled post-2022
multi-author peer-reviewed papers (cohort C).
Both cohorts are stratified random samples of 5{,}000 papers per field across
the 26 fields, plus 5{,}000 papers per subfield for the twenty subfields with
the largest trend breaks, drawn year-proportionally; pre-2022 collaborative
papers come from an identically drawn 2018--2021 multi-author sample
(cohort D).
Authors with more than 2{,}000 lifetime works or more than 50 works per
active career year are excluded as likely institutional placeholder profiles.

This construction follows from the estimand: we compare the content of solo
and coauthored output, not two disjoint populations, so an author who
publishes both types contributes to both sides.
Restricting to ``switchers'' who abandoned coauthoring would answer a
different question, force an arbitrary treatment of authors who do both, and
leave selection intact, since switching is itself a choice.

The groups are compared by regression adjustment alone: each outcome is
regressed on a solo-writer indicator with field fixed effects (the modal
field among the sampled papers underlying each observation) and, as
controls, the number of sampled post-2022 papers underlying the observation
and the number of the author's sampled pre-2022 papers, with
heteroskedasticity-robust standard errors.
The group labels are shorthand for sampled-paper authorship, not for author
histories: solo-writers are not required to be first-time solo authors, and
956 authors (15\% of the 6{,}521 solo-writers entering at least one of the
two outcome analyses) appear in both groups because the samples contain both
a solo and a coauthored paper by them.
For an author appearing in both groups, each outcome is computed separately
within each group over the corresponding papers: the solo-writer observation
uses the author's sampled solo papers and the persistent-coauthor observation
uses their sampled coauthored papers, so such an author contributes one
observation to each group.

Outcomes are computed over each author's sampled papers of the corresponding
authorship type.
Content breadth is the mean pairwise cosine distance among an author's
post-2022 papers in the corresponding group, computable for authors with at
least two sampled papers in that group (capped at twelve randomly chosen
papers per author): 0.089 for solo-writers vs.\ 0.115 for persistent
coauthors, 23\% narrower (regression-adjusted difference $-0.028$,
$P < 10^{-16}$; $n = 35{,}893$ authors: 3{,}342 solo-writers and 32{,}551
persistent coauthors).
Exploration is the mean cosine distance of an author's post-2022 papers in
the corresponding group from the centroid of their own pre-2022 coauthored
papers, computable for authors with at least one sampled pre-2022 paper; it
is statistically indistinguishable between the groups ($+0.001$, $P = 0.07$;
$n = 45{,}642$ authors: 3{,}494 and 42{,}148).
This comparison is therefore a covariate-adjusted descriptive contrast, not a
matched design, and it does not rule out residual selection into going solo;
we note, however, that the exploration outcome itself speaks against the most
natural selection story, in which authors go solo because their interests
have already moved away from their collaborative work.

\begin{figure*}[t]
  \centering
  \includegraphics[width=0.85\linewidth]{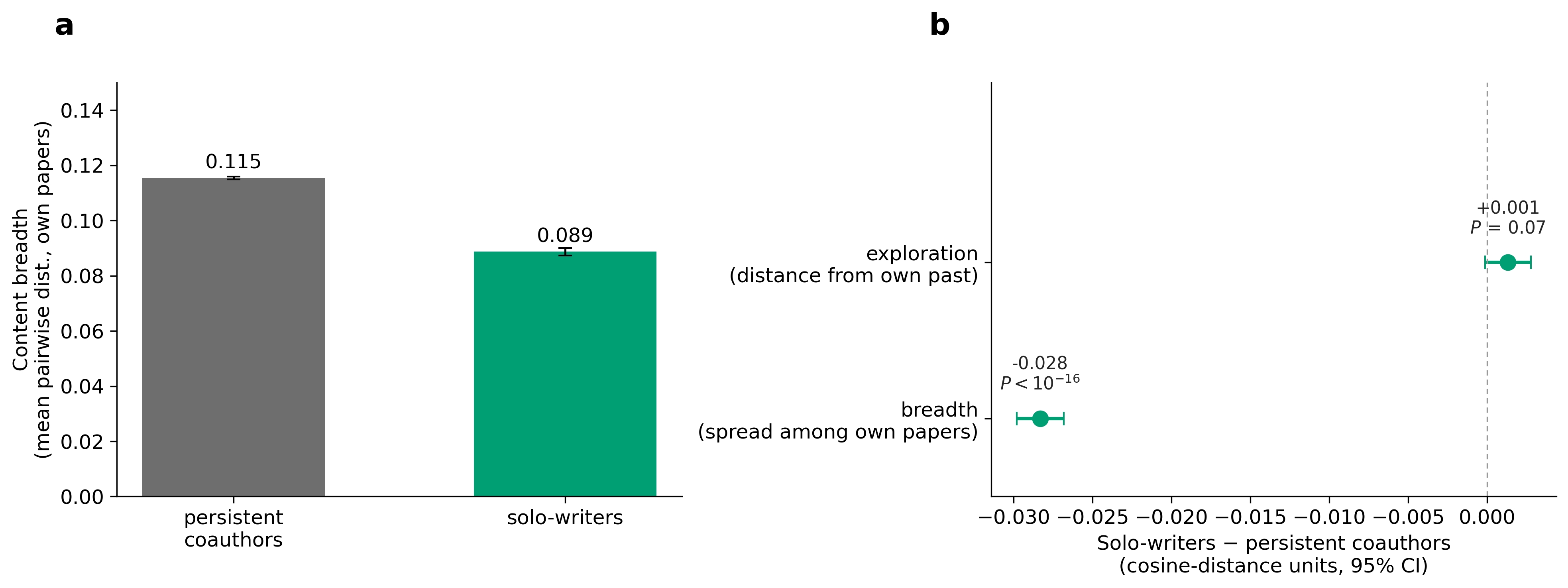}
  \caption{\textbf{Within authors, going solo contracts content scope.}
    Author-level comparison of solo-writers with persistent coauthors
    (Section~J).
    \textbf{a}, Content breadth: mean pairwise cosine distance among
    an author's own post-2022 papers of the corresponding authorship type
    (solo papers for solo-writers, coauthored papers for persistent
    coauthors). Bars, group means; error bars, 95\%
    confidence interval (CI) of the mean.
    \textbf{b}, Regression-adjusted differences (solo-writers $-$
    persistent coauthors) for breadth (spread among own post-2022
    papers) and for exploration (distance from the author's own
    pre-2022 collaborative centroid). Dots, point estimates; error bars, 95\%
    CI (two-sided $t$-tests).}
  \label{fig:author-focus}
\end{figure*}
\clearpage
\section{A complementary check using external data}
\label{sec:llm_modification_by_author_count}

A complementary check using external data points in the same direction.
Re-estimating the population-level LLM-modification rate of Liang et al.~\cite{liang2025quantifying} by author count, multi-author papers carry at least as much LLM-modified writing as solo papers in every arXiv field, and the rate is steepest for corresponding authors in non-native-English settings (SI Fig.~S10).
Writing assistance is therefore pervasive rather than concentrated in the solo tail, so surface-level editing cannot by itself account for a shift specific to solo authorship.

The panels of SI Fig.~S10 re-estimate the fraction of LLM-modified text ($\alpha$) in the public corpus of Liang et al.~\cite{liang2025quantifying} (arXiv, bioRxiv, and \emph{Nature} portfolio abstracts), using their population-level distributional quantification framework: for each field, reference word-frequency distributions for human- and LLM-written text parameterize a two-component mixture, and $\alpha$ is the maximum-likelihood fraction of LLM-modified sentences in a target corpus, estimated at the population level without classifying individual documents.
We re-bag the corpus by author count while holding the estimator and the reference distributions fixed.
Author counts are parsed from author strings (comma-delimited on arXiv, semicolon-delimited on bioRxiv); \emph{solo} denotes a solo author and \emph{multi} two or more.
Estimates pool all months within each window, with \emph{pre}~$=2021$ and \emph{post}~$=2024$ (January--September), and 95\% confidence intervals are obtained from $1{,}000$ sentence-level bootstrap resamples.
As a validity check, whole-venue estimates reproduce the published field-level values (e.g., computer science $\sim 22.5\%$, mathematics $\sim 7.8\%$, and the \emph{Nature} portfolio $\sim 8.9\%$ for September~2024).

In every arXiv field, multi-author papers have higher $\alpha$ than solo-authored papers ($+1.7$ to $+3.7$ percentage points, with largely non-overlapping intervals); bioRxiv is the only reversal, but it is based on a small, atypical solo subsample ($n \sim 300$).
Because $\alpha$ measures text-level modification rather than substitution for a collaborator's contribution, the consistent \emph{multi}~$\geq$~\emph{solo} pattern indicates that the reversal in the solo-authorship share (Fig.~1) is not explained by solo authors adopting writing assistance more heavily; the regional gradient (highest for China) is instead consistent with English-language polishing by non-native speakers, who are more likely to appear on larger teams.
Because $\alpha$ is a population-level measure of surface text modification, it does not capture author-level adoption, undetected drafting or analysis assistance, or differential effects of similar assistance on the decision to publish alone; we therefore read this check narrowly, as evidence against a surface-editing account.
This panel set uses an external corpus and a different outcome than our main analysis and is intended as a mechanism check complementing the author-count trends in SI Fig.~S2.

\begin{figure*}[t!]
\centering
\includegraphics[width=0.8\textwidth]{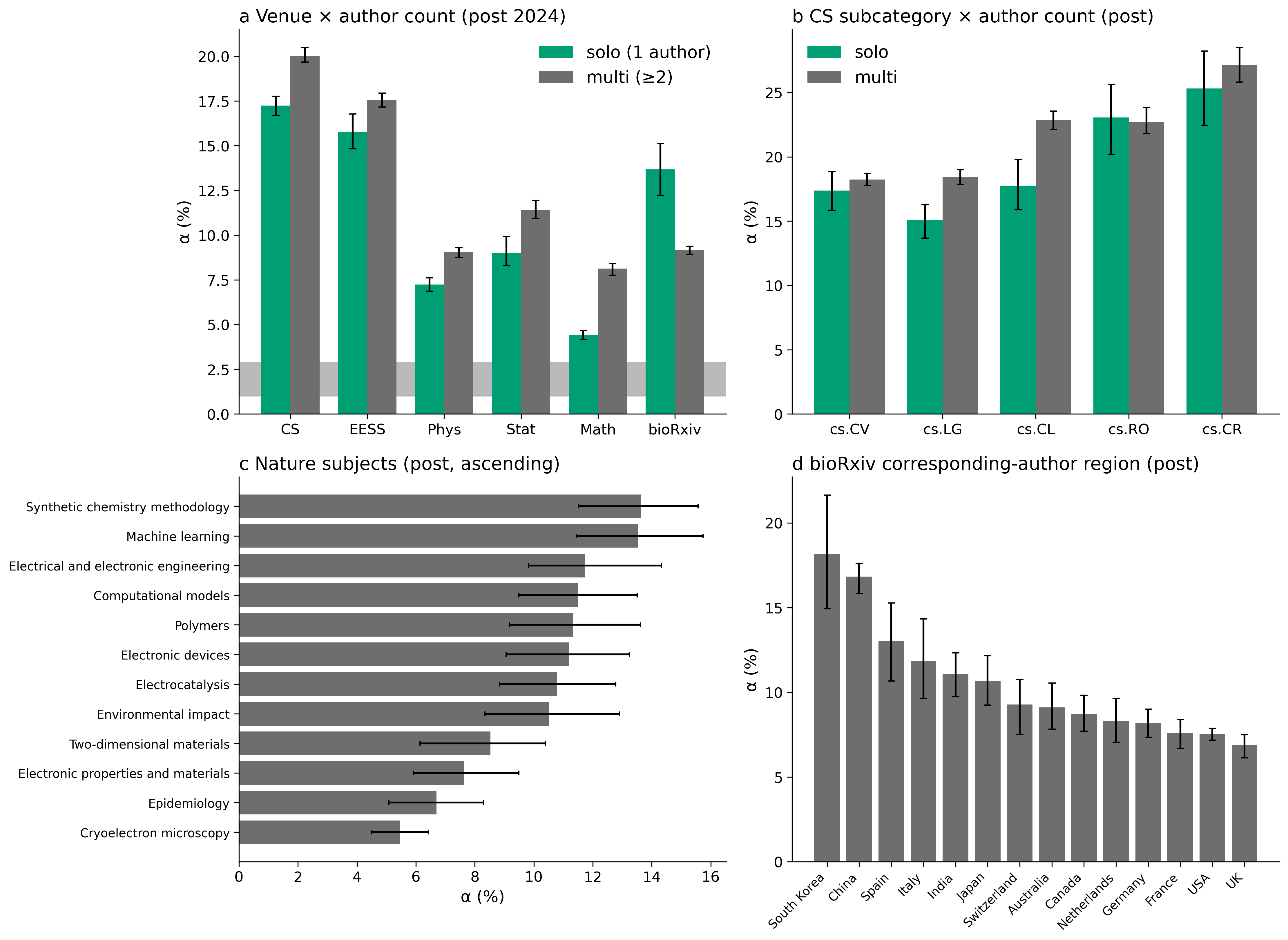}
    \caption{LLM-modified writing does not track solo authorship: after the
    public release of ChatGPT, multi-author papers show at least as much
    estimated LLM modification as solo-authored papers in every arXiv field;
    bioRxiv, based on a small, atypical solo subsample, is the only exception
    (estimation details in Section~K).
    \textbf{a},~Solo versus multi-author $\alpha$ (post) across five arXiv fields
    and bioRxiv; the grey band marks the flat pre-ChatGPT baseline (2021,
    $\sim 2$--$3\%$ in all venues).
    \textbf{b},~The same contrast within the five largest computer-science
    subcategories, showing that the solo--multi gap is not an artifact of subfield
    composition. \textbf{c},~The \emph{Nature} portfolio disaggregated by subject
    tag, where the aggregate ($\sim 8.9\%$) conceals a $5.5$--$13.6\%$ spread.
    \textbf{d},~bioRxiv $\alpha$ by corresponding-author region; region is
    assigned by a keyword heuristic that classifies only $\sim 20\%$ of
    institutions and is shown as illustrative, with a ROR-based
    affiliation--country mapping preferable.}
\label{fig:llm_modification_by_author_count}
\end{figure*}

\clearpage
\section{The trend break within continuously observed venues}
\label{sec:si_balanced_venues}

OpenAlex's coverage and record-processing systems changed as the database was developed and updated, including the transition from Microsoft Academic Graph after MAG was discontinued at the end of 2021~\cite{openalex_mag_replacement} and a substantial revision of OpenAlex's author-disambiguation system in July 2023~\cite{openalex_author_disambiguation}. Such database-specific idiosyncrasies therefore require careful attention~\cite{alperin2024analysis, culbert2025reference}.

This creates two distinct risks.
First, venues may enter or leave the OpenAlex index near the panel's right edge, changing the aggregate solo-authorship share through composition alone.
Second, authorship metadata or record processing may change within venues that remain continuously indexed.
The analysis below partially addresses the first risk but cannot eliminate the second.

To assess the venue-composition margin, we re-estimate the trend break on a balanced venue panel: for each filter variant, we retain host sources, defined as the source of a work's primary location, with at least one qualifying paper in every calendar year from 2018 to 2024.
The year 2025 is exempt from this requirement because of indexing lag, but 2025 papers from qualifying sources enter the estimation; works without a source identifier drop out by construction.
The balanced peer-reviewed panel comprises 45{,}621 sources carrying 72--77\% of peer-reviewed papers in every year, and the balanced core panel comprises 19{,}616 sources carrying 92--95\% of core papers.
A stricter variant additionally requires the source to be first observed in 2015 or earlier, yielding 36{,}243 sources carrying 65--70\% of peer-reviewed papers.

The break persists within continuously observed venues at roughly half its full-sample magnitude.
The pooled monthly piecewise fit gives $\Delta\beta = +0.75$~pp/yr (95\% CI $[+0.41, +1.10]$) on the balanced peer-reviewed panel and $+0.82$ $[+0.44, +1.20]$ on the stricter panel, compared with $+1.72$ $[+1.12, +2.32]$ in the full sample (SI Fig.~S11a).
The corresponding core estimates, not shown in the figure, are $+0.53$ $[+0.25, +0.82]$ in the balanced panel and $+0.70$ $[+0.39, +1.01]$ in the full sample.
Within balanced venues, the pre-2022 decline of $-1.18$~pp/yr flattens to $-0.42$~pp/yr after the break.
The attenuation relative to the full sample suggests that venue composition contributes roughly half of the pooled break, while a positive and statistically significant break remains in the balanced-panel specifications.
The cross-field pattern is preserved: the Spearman rank correlation between full-sample and balanced-panel $\Delta\beta_{2022}$ across the 26 fields is $0.84$ for the peer-reviewed filter and $0.91$ for the core filter (SI Fig.~S11b).

Two deviations are informative.
Engineering attenuates from $+2.47$ to $+0.65$ $[+0.60, +0.70]$, consistent with the compositional interpretation of its field-level break (Section~H) and indicating that part of this composition operates through the venue margin.
Arts and Humanities flips from $-0.50$ to $+0.16$ $[+0.02, +0.31]$, suggesting that its full-sample decline partly reflects venue composition rather than changes within continuously observed venues.
Materials Science under the core filter moves similarly but is statistically flat in both samples ($+0.10$ to $-0.02$).

\begin{figure*}[t]
  \centering
  \includegraphics[width=0.85\linewidth]{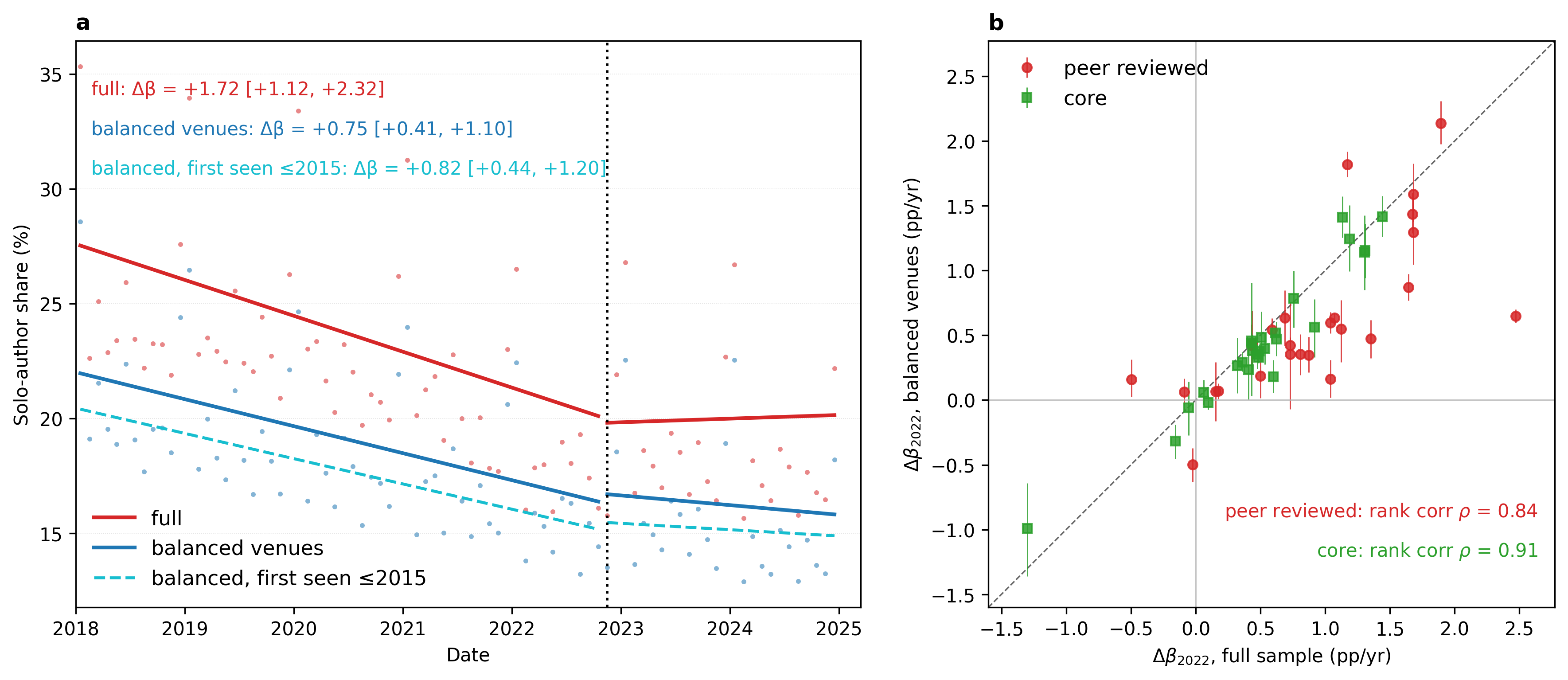}
  \caption{The 2022 trend break within continuously observed venues (Section~L).
    \textbf{a}, Pooled monthly solo-author share under the peer-reviewed filter: full sample, the balanced panel of sources publishing in every year 2018--2024, and the stricter balanced panel additionally requiring first observation by 2015 (dashed); lines are separate weighted least-squares fits before and after November 2022 (calendar-month effects removed; 2025 excluded from the fit window), with $\Delta\beta$ and 95\% CIs annotated.
    \textbf{b}, Field-level $\Delta\beta_{2022}$ (symmetric four-year windows, Section~B) in the balanced panel against the full sample, for the peer-reviewed and core variants, with Spearman rank correlations annotated; error bars, binomial bootstrap 95\% confidence intervals; the dashed line marks equality.}
  \label{fig:si_balanced}
\end{figure*}

\clearpage
\section{Timing robustness of the 2022 break}
\label{sec:si_timing}

Three checks address the timing of the break directly.
First, an event study: the field-year solo share is regressed on calendar-year indicators (reference year 2018) with field fixed effects, weighted by paper counts, with standard errors clustered by field (SI Fig.~S12a).
Under the main peer-reviewed specification the share declines monotonically through 2022, reaching $-7.2$~pp relative to 2018 with the year-over-year change steepening to $-2.9$~pp in 2022; the decline then halts abruptly ($+0.1$ in 2023, $-0.6$ in 2024, $-0.3$ in 2025), and no earlier year shows a flattening of comparable size.
The pre-trend is deliberately left visible: the claim concerns the change in trend, not the level.
Second, donut specifications remove the transition window entirely.
Excluding 2022 from both windows, the annual pooled $\Delta\beta$ is $+0.82$ $[+0.79, +0.85]$ under peer reviewed ($+1.09$ with 2022 opening the post window).
Excluding November 2022 through June 2023 from the monthly fit leaves the point estimate close to the published one ($+1.38$ versus $+1.72$), with a mechanically wider Newey--West confidence interval $[-0.03, +2.79]$, as the donut removes eight of the twenty-six post months.
Third, 2025 is excluded: with the post window limited to 2023--2024, the annual pooled $\Delta\beta$ is $+0.78$ $[+0.73, +0.83]$ under peer reviewed and $+0.22$ $[+0.18, +0.27]$ under core; the published monthly fits already end in December 2024 and are unaffected by construction.
One caveat cuts the other way: under the unrestricted \emph{all} coverage, the post-2023--24 estimate turns negative ($-1.50$), that is, the raw-all break depends on provisional 2025 data, an additional reason the paper's headline variants are the venue-restricted ones (Section~A).

\begin{figure*}[t]
  \centering
  \includegraphics[width=0.85\linewidth]{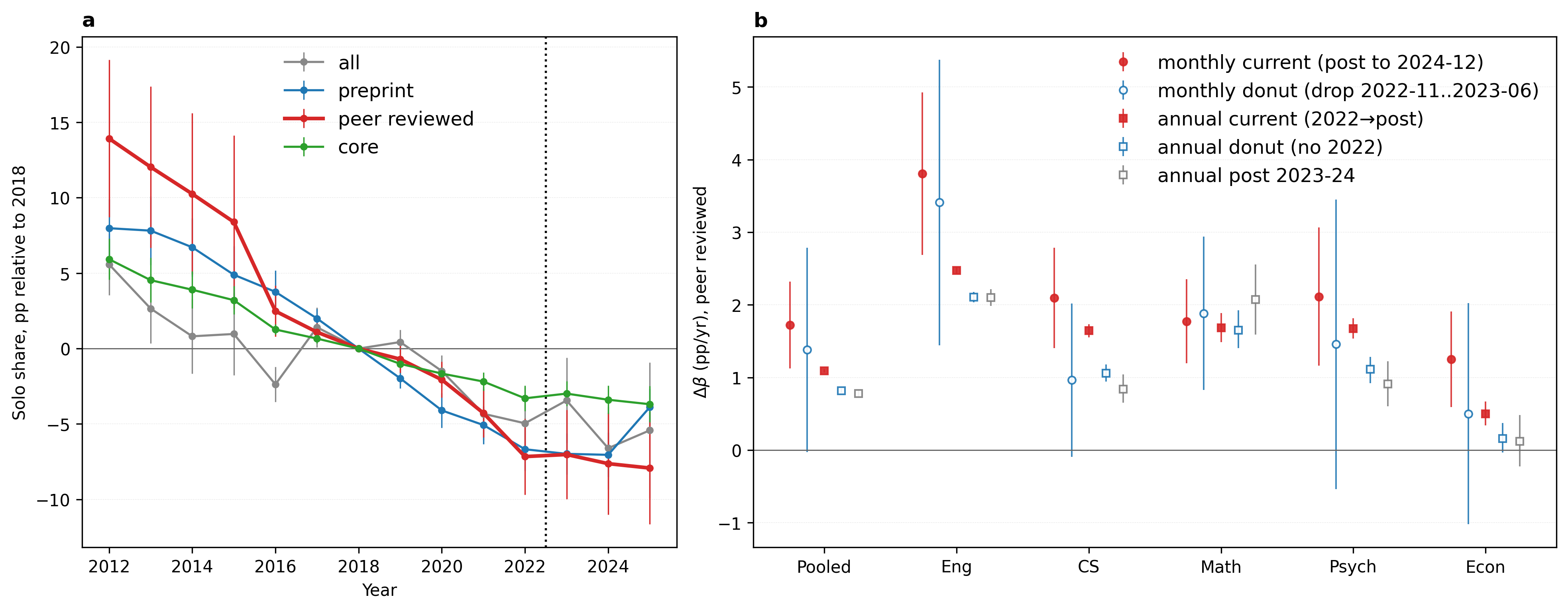}
  \caption{Timing robustness of the 2022 break (Section~M).
    \textbf{a}, Event-study coefficients: field-year solo share regressed on calendar-year indicators (reference year 2018) with field fixed effects and paper-count weights, by filter variant; whiskers are 95\% confidence intervals with field-clustered standard errors; the dotted line marks the public release of ChatGPT.
    \textbf{b}, Pooled and focal-field $\Delta\beta$ under the peer-reviewed variant across window specifications: the published monthly fit (post window ending December 2024), a monthly donut dropping November 2022--June 2023, the annual specification with 2022 opening the post window, an annual donut excluding 2022, and an annual specification with the post window limited to 2023--2024; whiskers are 95\% confidence intervals (Newey--West for monthly, binomial bootstrap for annual). ``Pooled’’ denotes the estimate on the base sample with all fields combined.}
  \label{fig:si_timing}
\end{figure*}

\end{document}